\documentclass[twocolumn]{aastex62}

\received{Sept 24, 2019}
\published{January 24, 2020}

\begin{document}

\newcommand{\kms}{km~s$^{-1}$} \newcommand{\msun}{$M_{\odot}$} 
\newcommand{\rsun}{$R_{\odot}$} \newcommand{\teff}{$T_{\rm eff}$} 
\newcommand{\logg}{$\log{g}$} \newcommand{\mas}{mas~yr$^{-1}$}

\title{ The ELM Survey. VIII.  98 Double White Dwarf Binaries}

\author{Warren R.\ Brown} \affiliation{Smithsonian Astrophysical Observatory, 60 
Garden Street, Cambridge, MA 02138 USA}

\author{Mukremin Kilic} \affiliation{Homer L. Dodge Department of Physics and 
Astronomy, University of Oklahoma, 440 W. Brooks St., Norman, OK, 73019 USA}

\author{Alekzander Kosakowski} \affiliation{Homer L. Dodge Department of Physics and 
Astronomy, University of Oklahoma, 440 W. Brooks St., Norman, OK, 73019 USA}

\author{Jeff J.\ Andrews} \affiliation{Niels Bohr Institute, Blegdamsvej 17, DK-2100 
Kopenhagen, Denmark}

\author{Craig O.\ Heinke} \affiliation{Department of Physics, University of Alberta, 
Edmonton, AB T6G 2E1, Canada}
 
\author{Marcel A. Ag{\"u}eros} \affiliation{Department of Astronomy, Columbia University,
550 West 120th Street, New York, NY 10027, USA}

\author{Fernando Camilo} \affiliation{South African Radio Astronomy Observatory, 2 
Fir Street, Observatory 7925, South Africa}

\author{A.\ Gianninas} \affiliation{Department of Physics and Astronomy, Amherst College,
25 East Drive, Amherst, MA 01002, USA} 

\author{J.J.\ Hermes} \affiliation{Department of Astronomy, Boston University, 725 
Commonwealth Ave., Boston, MA 02215, USA}

\author{Scott J.\ Kenyon} \affiliation{Smithsonian Astrophysical Observatory, 60 
Garden Street, Cambridge, MA 02138 USA}

\email{wbrown@cfa.harvard.edu}

\shorttitle{ The ELM Survey. VIII.  98 White Dwarf Binaries } 
\shortauthors{Brown et al.}

\begin{abstract}

	We present the final sample of 98 detached double white dwarf (WD) binaries 
found in the Extremely Low Mass (ELM) Survey, a spectroscopic survey targeting 
$<$0.3 \msun\ He-core WDs completed in the Sloan Digital Sky Survey footprint.  
Over the course of the survey we observed ancillary low mass WD candidates like 
GD~278, which we show is a $P=0.19$~d double WD binary, as well as candidates 
that turn out to be field blue straggler/subdwarf A-type stars with luminosities too 
large to be WDs given their {\it Gaia} parallaxes.  Here, we define a clean sample 
of ELM WDs that is complete within our target selection and magnitude range 
$15<g_0<20$ mag.  The measurements are consistent with 100\% of ELM WDs being 
$0.0089 < P < 1.5$~d double WD binaries, 35\% of which belong to the Galactic halo.  
We infer these are mostly He+CO WD binaries given the measurement constraints.  The 
merger rate of the observed He+CO WD binaries exceeds the formation rate of stable 
mass transfer AM CVn binaries by a factor of 25, and so the majority of He+CO WD 
binaries must experience unstable mass transfer and merge.  The shortest-period 
systems like J0651+2844 are signature {\it LISA} verification binaries that can be 
studied with gravitational waves and light.

\end{abstract}


\section{INTRODUCTION}

	The Milky Way is expected to contain $\mathcal{O}(10^8)$ double degenerate 
white dwarf (WD) binaries \citep[][]{han98, nelemans01a} because most stars evolve 
into WDs and most stars reside in binaries \citep[][]{moe17}.  Ultra-compact WD 
binaries, with orbital periods of hours to minutes, are particularly interesting 
because they are strong mHz gravitational wave sources that will be detected by the 
{\it Laser Interferometer Space Antenna} ({\it LISA}) \citep{amaro17}.  
Gravitational wave radiation causes ultra-compact WD binaries to lose orbital 
energy, eventually turning into stable mass transfer AM~CVn systems, supernovae, or 
single massive WDs, R~CrB stars, and related objects \citep[e.g.][]{webbink84}.

	Theoretical models have long predicted that most ultra-compact WD binaries 
contain low mass, He-core WDs \citep[e.g.][]{iben90}.  Observationally, this means 
low mass WDs are the signposts of ultra-compact binaries \citep{iben97b}.  Indeed, 
when \citet{marsh95} observed seven of the lowest mass 0.3 -- 0.4~\msun\ WDs in the 
\citet{mccook87} catalog, they found five WD binaries with orbital periods of 
$P=4$~h to 4~d.  By comparison, the ESO Supernovae type Ia Progenitor Survey 
targeted 643 normal hydrogen-atmosphere WDs and found 39 binaries, the majority of 
which are the lowest mass WDs in their sample \citep{napiwotzki19}.

	Here we present the completed ELM Survey, a spectroscopic survey that 
targeted ``extremely low mass'' (ELM) $<$0.3 \msun\ WDs in the Sloan Digital 
Sky Survey (SDSS) footprint \citep{kilic10, kilic11a, brown10c, brown12a}.  We 
refer to WDs with $5\lesssim \log{g} \lesssim 7$ as ELM WDs because they are 
essentially absent in other WD catalogs \citep{eisenstein06, gianninas11, 
kleinman13, napiwotzki19} that targeted normal $\log{g}=8$ WDs.  We have used 
previous versions of our sample to address the space density, orbital distribution, 
and merger rate for this class of double WD binary \citep{gianninas15, brown16a, 
brown16b}.

	The completed ELM Survey contains over half of the known detached double WDs 
in the Galaxy \citep{marsh19}.  Our approach was to target candidates that have 
the magnitudes and colors of a single low mass WD.  We find that most of the low 
mass WDs are single-lined spectroscopic binaries; the companions are significantly 
fainter than the observed low mass WD by survey design.  Our approach is thus a 
productive way of finding double degenerate binaries.  The results inspired us to 
search for $\log{g}\sim6$ objects in other spectroscopic catalogs.  We include a few 
dozen additional low mass WD candidates we found in other spectroscopic catalogs to 
the final ELM Survey sample published here.

	In Section 2, we present the 4,338 radial velocity measurements and 230 
stellar atmosphere fits for the completed ELM Survey sample.  We apply {\it Gaia} 
parallax and proper motion measurements to the sample for the first time.  We find 
that stellar atmosphere-derived luminosity estimates are in excellent agreement with 
{\it Gaia} parallax measurements at effective temperatures $T_{\rm eff}>9,$500~K.  
However, many of the coolest $<$9,000~K objects, where the hydrogen Balmer lines 
lose their sensitivity to temperature and gravity \citep{strom69}, are subdwarf 
A-type stars \citep{kepler15, kepler16}.  {\it Gaia} parallax shows that most 
subdwarf A-type stars are mis-identified metal poor halo stars \citep{brown17a, 
pelisoli17, pelisoli18b, pelisoli18a, pelisoli19a, yu19}; at these temperatures, 
such stars are also called field blue stragglers \citep[e.g.][]{bond71, preston00}.  
The focus of this paper is WDs, and so we define clean samples that exclude all 
non-WD stars.

	In Section 3, we present radial velocity orbital parameters for the 
full set of 128 binaries in the ELM Survey, including 25 well-constrained new
systems. We use previously unpublished optical, radio, and X-ray observations to 
provide inclination constraints for 47 of the binaries.  In Section 4, we study the 
distribution of WD binary properties and compare their gravitational wave 
strain to {\it LISA} sensitivity curves. We conclude in Section 5.

\section{DATA}

	In this section, we consolidate the measurements from the full ELM 
Survey and publish the final set of discoveries.  We observed a total of 230 low 
mass WD candidates with $>3$ spectroscopic observations.  {\it Gaia} parallax shows 
that the 230 candidates are a mixed bag of objects, and so we close this 
Section by defining a clean ELM WD sample.

\subsection{Target Selection}

	We select low mass WD candidates for the ELM Survey on the basis of 
broadband color using SDSS photometry \citep{alam15}.  The first targets were found 
serendipitously in the MMT Hypervelocity Star Survey \citep{brown06b, kilic07}.  We 
then designed the ELM Survey to find more low mass WDs.

	The color selection is detailed in previous HVS Survey \citep{brown12b} and 
ELM Survey papers \citep{brown12a}.  We select using de-reddened magnitudes and 
colors, indicated by the subscript 0.  Having $u$-band is the key to the target 
selection.  Physically, the $(u-g)_0$ color spans the Balmer decrement and provides 
a sensitive measure of surface gravity at 10,000 -- 20,000~K temperatures or $-0.4 < 
(g-r)_0 < -0.1$ mag colors.  We used $(r-i)_0$ to exclude sources with non-stellar 
colors such as quasars.

	We also select a few dozen low mass WD candidates from pre-existing 
catalogs:  every object we could find listed with $\log{g}\sim6$.  We found 
most of the additional candidates in the SDSS spectroscopic catalog.  However, we 
also found one candidate (WD0921$-$120 = J0923$-$1218) in the Edinburgh-Cape Survey 
\citep{kilkenny97}, one candidate (GD278 = J0130+5321) in the {\it TESS} bright WD 
catalog \citep{raddi17}, and two candidates (J0308+5140 and J1249+2626) in the 
LAMOST catalog \citep{lamost15}.  The candidates are outliers in their 
catalogs, and do not represent a complete sample in any way.  Indeed, the additional 
candidates turn out to be a diverse set of objects including mis-identified
hot subdwarf B stars, cooler subdwarf A stars, as well as some ELM WDs.  The 
additional candidates provide a useful context to the main ELM Survey, but  
emphasize the importance of follow-up spectroscopy.

\subsection{Survey Design}

	Our approach is to acquire a single spectrum for every candidate and 
determine its nature using stellar atmosphere fits.  The observations are 99\% 
complete for all candidates in the range $15<g_0<20$ mag.

	We then acquire multi-epoch spectroscopy for candidates that appear to be 
$5<\log{g}<7$ WDs.  Our multi-epoch observations are 97\% complete.  Some objects 
inevitably fall outside our primary $\log{g}$ selection upon further observation;
however, we continue to observe any candidate showing radial velocity variability.  

	The upshot is that our sample of {\it binaries} is effectively selected on 
the basis of magnitude, color (temperature), and surface gravity; we deliberately 
re-observed all $5<\log{g}<7$ WD candidates.  However the color selection does not 
evenly sample \logg\ at all temperatures (see Figure 1 and also \citealt{brown12a}).  
In practice, our sample of binaries contains objects up to $\log{g}\sim7.5$ at 
$T_{\rm eff}>12,000$~K and an over-abundance of $\log{g}<6$ objects at $T_{\rm 
eff}<9,000$~K.  We discuss a clean sample in Section 2.7.
	There are a total of 230 candidates with $>$3 epochs of observations.

\subsection{Spectroscopy}

	The low mass WD candidates, given our color-selection, have A-type 
spectra dominated by hydrogen Balmer lines.  The high-order Balmer lines are 
sensitive to surface gravity \citep{tremblay09} and provide a good measure of radial 
velocity.  Thus we acquire spectra using spectrographs with good near ultra-violet 
sensitivity.

	Most of our spectroscopy was obtained with the 6.5m MMT telescope and the 
Blue Channel spectrograph.  Stellar atmosphere fits, with few exceptions, are done 
with MMT spectra.  The majority of binary orbits are also derived from time-series 
MMT spectra, though we acquired time-series spectra for the brightest $g<17$ mag 
objects at the 1.5m Fred Lawrence Whipple Observatory (FLWO) telescope and, starting 
in 2017, the 4.1m Southern Astrophysical Research (SOAR) telescope.  For purposes of 
completeness, we also include radial velocity measurements acquired at the 8m Gemini 
telescopes and the 4m Mayall telescope.

	At the 6.5m MMT telescope, we acquire spectra using the Blue Channel 
spectrograph \citep{schmidt89} with the 832 l~mm$^{-1}$ grating in 2nd order and a 
1.0 or 1.25 arcsec slit.  This set-up provides us with 1.0 or 1.2~\AA\ spectral 
resolution over $3550 < \lambda < 4500$~\AA.  We normally set exposure times to 
yield signal-to-noise (S/N) $\sim$7 per pixel, or S/N $\sim$12 per resolution 
element, per exposure.  The exception to this rule were the short $P<40$ min 
binaries, which we additionally observed with the 800 l~mm$^{-1}$ grating in 1st 
order and a 1.0 arcsec slit.  The lower throughput of this set-up is offset by the 
2.4~\AA\ spectral resolution and greater $3550 < \lambda < 5500$ \AA\ spectral 
coverage, enabling shorter exposure times and better time resolution of the shortest 
orbital period binaries.

\begin{deluxetable}{ccc}	
 \tablecolumns{3} \tablewidth{0pt} 
 \tablecaption{Radial Velocity Measurements\label{tab:rv}} 
 \tablehead{
        \colhead{Object}& \colhead{HJD} & \colhead{$v_{helio}$}\\
                        & -2450000 (days) & (\kms ) }
        \startdata 
0027-1516  &  5385.970567  & $  90.47 \pm  9.29$ \\
0027-1516  &  6126.921060  & $  89.67 \pm  8.66$ \\
0027-1516  &  6126.933677  & $  83.02 \pm 14.25$ \\
0027-1516  &  6126.955646  & $  53.73 \pm 11.97$ \\
0027-1516  &  6244.709757  & $-160.29 \pm 17.63$ \\
0027-1516  &  6244.742092  & $-182.75 \pm 10.36$ \\
0027-1516  &  7008.557845  & $-212.40 \pm  9.83$ \\
0027-1516  &  7008.606370  & $ -91.13 \pm 13.60$ \\
0027-1516  &  7012.560040  & $ 135.84 \pm  8.77$ \\
0027-1516  &  7012.580906  & $ 101.15 \pm  8.48$ \\
0027-1516  &  7012.647103  & $  -9.19 \pm 12.61$ \\
0027-1516  &  7336.570378  & $  66.47 \pm 19.86$ \\
0027-1516  &  7336.669966  & $-136.06 \pm 16.48$ \\
0027-1516  &  7337.769135  & $ 162.85 \pm 28.61$ \\
0027-1516  &  7359.557542  & $ -38.19 \pm 14.93$ \\
0027-1516  &  7723.561348  & $-172.08 \pm 16.22$ \\
0027-1516  &  7723.711867  & $ 108.65 \pm 13.93$ \\
        \enddata
 \tablecomments{This table is available in its entirety in machine-readable and
	Virtual Observatory forms in the online journal. A portion is shown here
	for guidance regarding its form and content.}
 \end{deluxetable}

	At the 1.5m FLWO telescope, we acquire spectra using the FAST spectrograph 
\citep{fabricant98} with the 600 l~mm$^{-1}$ grating and the 1.5 arcsec slit.  This 
set-up provides 1.7~\AA\ spectral resolution over $3550 < \lambda < 5500$~\AA.  We 
normally set exposure times to yield S/N $\sim$15 per pixel, or S/N $\sim$23 per 
resolution element, per exposure, to compensate for the lower spectral resolution 
compared to the MMT telescope.

	At the 4.1m SOAR telescope, we acquire spectra using the Goodman High 
Throughput spectrograph \citep{clemens04} with the 930 l~mm$^{-1}$ grating and the 
1.03 arcsec slit.  This set-up provides 2.2~\AA\ spectral resolution over $3550 < 
\lambda < 5250$ \AA.  The SOAR spectra were obtained as part of the NOAO program 
2017A-0076.

	At the 8m Gemini telescopes, we acquire spectra using the Gemini Multi 
Object Spectrograph \citep{hook04} with the B600 grating and the 0.5 arcsec slit 
\citep{kilic17}.  This set-up provides 2.1 \AA\ spectral resolution over $3600 < 
\lambda < 6600$ \AA.

	At the 4m Mayall telescope, we acquire spectra using the Kitt Peak Ohio 
State Multi-Object Spectrograph \citep{martini14} using the Blue VPH grating and the 
1.5 arcsec slit.  This set-up provides 2.0 \AA\ spectral resolution over $3500 < 
\lambda < 6200$ \AA.  Throughput and calibration below 4000 \AA\ is poor compared to 
the other spectrographs, and very little 4m Mayall data are used. The Kitt Peak 
spectra were obtained as part of the NOAO program 2016B-0160.

	We paired all observations with a comparison lamp exposure for accurate 
wavelength calibration, and measured radial velocities with the cross-correlation 
package RVSAO \citep{kurtz98} using high S/N templates obtained with the same 
spectrograph set-up.  We use the full wavelength range of the spectra, which 
typically contain 6 to 10 well-measured Balmer lines depending on the target's 
surface gravity, to measure radial velocity. The median statistical velocity error 
is 15 \kms.  The systematic velocity zero-point error is 2-3 \kms\ based on a 
comparison of time-series spectra obtained for the same target at MMT, SOAR, and 
FLWO telescopes.

	We present 4,338 radial velocity measurements for 230 low mass WD candidates 
with $>3$ observations in Table~\ref{tab:rv}.  Two-thirds of the radial velocities 
in Table~\ref{tab:rv} are published in previous ELM Survey papers, and one-third 
have not been published before.  We consolidate everything into a single table for 
ease of use.  The new content in Table~\ref{tab:rv} includes observations for 25 
well-constrained new binaries, further observations for 30 of 99 previously 
published binaries, and observations for all other candidates with $>$3 epochs of 
observation.

\begin{figure} 
 \includegraphics[width=3.5in]{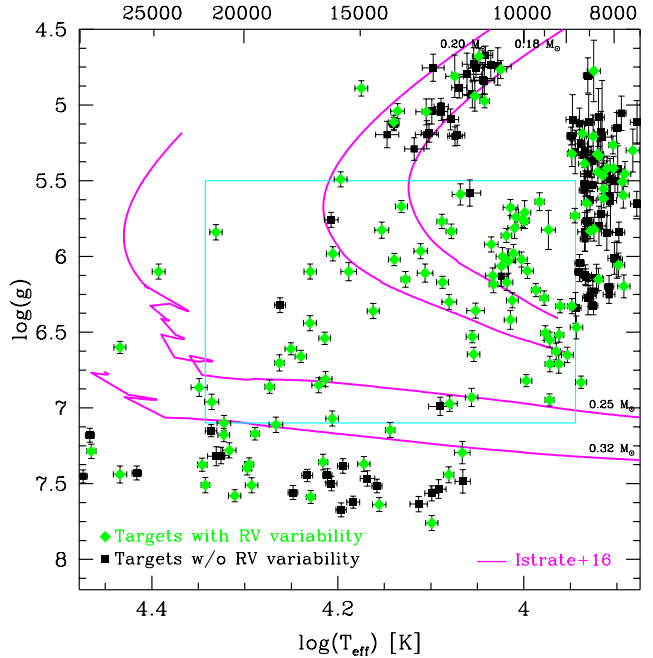} 
 \caption{ \label{fig:teff}
	Effective temperature \teff\ versus surface gravity \logg\ for all 230 
candidates with $>$3 spectroscopic observations.  Green points are candidates with 
significant radial velocity variability; the binaries.  Magenta lines are selected 
WD tracks for halo progenitors from \citet{istrate16}, with the shell flash loops in 
$>$0.25 \msun\ tracks clipped for the sake of clarity.  The cyan box marks the clean 
ELM WD sample, a region in which observations are complete in the range $15<g_0<20$ 
mag in the SDSS footprint. }
 \end{figure}

\subsection{Stellar Atmosphere Fits}

	We perform stellar atmosphere fits as described in previous ELM Survey 
papers.  We fit the summed, rest-frame spectra of each candidate to a grid of pure 
hydrogen atmosphere models that span 4,$000~{\rm K} < T_{\rm eff} < 35$,000~K and 
$4.5 < \log{g} < 9.5$ \citep{gianninas11, gianninas14b, gianninas15} and that 
include the Stark broadening profiles from \citet{tremblay09}.  We then apply the 
\citet{tremblay15} three-dimensional stellar atmosphere model corrections if needed.  
We present the corrected stellar atmosphere parameters for all 230 candidates in the 
electronic version of Table~\ref{tab:param}, but limit the print version of 
Table~\ref{tab:param} to the 25 well-constrained new binaries.

	Figure~\ref{fig:teff} plots the distribution of \teff\ and \logg\ for the 
230 candidates with multi-epoch observations.  The candidates with significant 
radial velocity variability -- the binaries -- are marked with green diamonds.  
Magenta lines are theoretical evolutionary tracks from \citet{istrate16} for halo 
($Z=0.001$ progenitor) WDs with masses ranging from 0.18 \msun\ to 0.32 
\msun.  We clip the loops due to shell flashes -- the discontinuities in the 
higher-mass tracks -- for purpose of illustration.

	The distribution of points in Figure~\ref{fig:teff} reflects our target 
selection convolved with our follow-up approach.  Our multi-epoch observations span 
candidates with $4.5<\log{g}<7.5$, but the follow-up is only complete for candidates 
with $5<\log{g}<7$.  The Survey contains many false-positives around 
$\log{g}\sim5$ because the underlying color selection pushes up against the locus of 
normal A-type stars -- field blue stragglers -- at low gravities.  At higher 
gravities, the color selection pushes up against the locus of normal DA-type WDs.

	In Figure~\ref{fig:teff}, candidates hotter than 12,000~K primarily come 
from the HVS Survey target selection.  Subdwarf B stars, objects found in the range 
25$,000 < T_{\rm eff} < 40,$000~K and $5<\log{g}<6$ \citep{heber09}, are 
deliberately excluded from the target selection (thus the empty upper left corner of 
Figure~\ref{fig:teff}).  Candidates cooler than 12,000~K primarily come from the ELM 
Survey target selection; the band of subdwarf A-type stars at $<$9,000~K is notable.  
The band of low-gravity objects around 12,000~K is also heavily contaminated by 
field blue stragglers (see discussion of {\it Gaia} results below).

 \begin{figure} 	
 \includegraphics[width=3.5in]{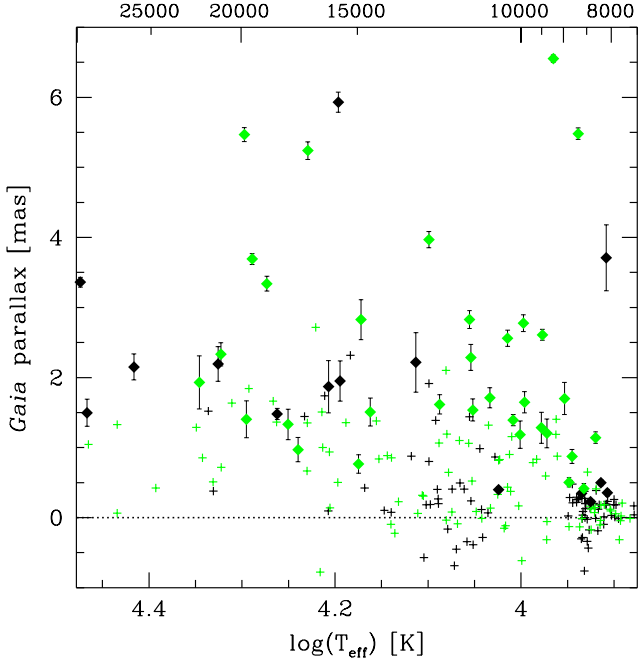}
 \caption{ \label{fig:plxteff}
	{\it Gaia} parallax vs.\ temperature, plotted on the same scale as 
Figure~\ref{fig:teff}.  Solid diamonds with errorbars mark candidates with
$\pi/\sigma_\pi>$5; plus signs mark everything else.  Green colors mark
binaries.  Dotted line marks zero parallax. } 
 \end{figure}

\subsection{{\it Gaia} Astrometry}

	We cross-match the 230 candidates against {\it Gaia} Data Release 2 (DR2) on 
the basis of position and apparent magnitude.  We find matches for all 230 
candidates, although 8 candidates lack 5-parameter (position, proper motion, and 
parallax) solutions.  Table~\ref{tab:param} presents the {\it Gaia} values for the 
sample.  For the eight objects without {\it Gaia} DR2 measurements, we present 
proper motions from {\it Gaia}-PanStarrs1 \citep{tian17}.

	Parallax provides a direct constraint on the stellar nature of the 
candidates. Figure~\ref{fig:plxteff} plots the distribution of parallax versus 
temperature for all 222 candidates with 5-parameter {\it Gaia} DR2 measurements. We 
apply the parallax zero-point offset 0.029~mas recommended by the {\it Gaia} team 
\citep{lindegren18}.  For the sake of clarity, we draw errorbars only for those 
candidates with parallax values greater than 5 times the parallax error, 
$\pi/\sigma_\pi>5$, the quality threshold used by the {\it Gaia} team 
\citep{lindegren18}.  The 53 candidates with $\pi/\sigma_\pi>5$ are marked as 
solid diamond in Figure \ref{fig:plxteff}; everything else is marked as a plus sign.

	Candidates with few mas parallaxes, or few hundred pc distances, are likely 
nearby WDs and are present at all temperatures in our sample.  Candidates with 
approximately zero parallax are much more distant and unlikely to be WDs.  The zero 
parallax objects are clumped around 12,000~K and 8,500~K, and correspond to the 
$\log{g}<5.5$ and $T_{\rm eff}<9,$000~K groups of candidates in Figure~\ref{fig:teff}.

 \begin{figure} 	
 \includegraphics[width=3.5in]{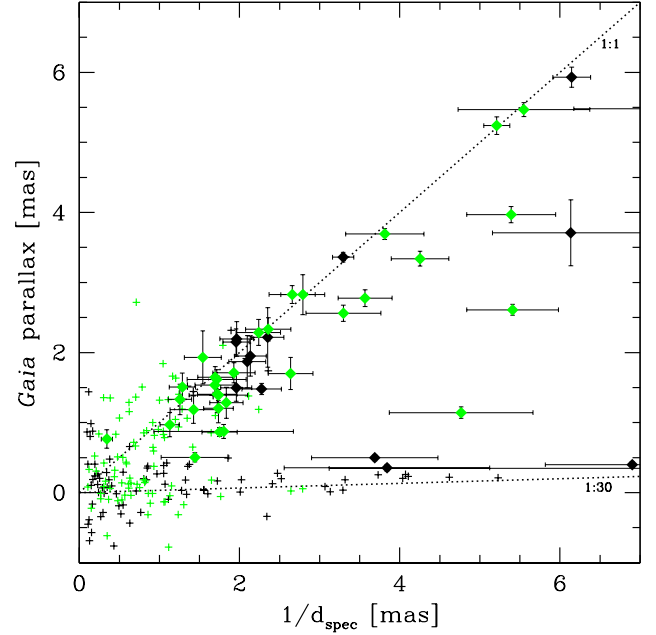}
 \caption{ \label{fig:plxplx}
	{\it Gaia} parallax vs.\ inverse spectrophotometric distance.  Symbols are 
the same as Figure \ref{fig:plxteff}. Dotted lines are the 1:1 and 1:30 parallax 
ratio lines.  The parallax distribution suggests that about half of the candidates 
are nearby WDs and half are distant subdwarf A (field blue straggler) stars.}
 \end{figure}

 \begin{splitdeluxetable*}{lcccccccccBccccccc}
   \tabletypesize{\scriptsize} \tablecolumns{17} \tablewidth{0pt} 
   \tablecaption{Measured and Derived Parameters\label{tab:param}} 
   \tablehead{
	\colhead{Object} & \colhead{R.A.} & \colhead{Decl.} & 
	\colhead{$g_0$} & \colhead{$T_{\rm eff}$} & \colhead{$\log g$} & 
	\colhead{WD} & \colhead{ELM} &  \colhead{Clean} & \colhead{Disk} &
	\colhead{Mass(WD=1)} & \colhead{$M_g$(WD=1)} & \colhead{$d_{helio}$(WD=1)} &
	\colhead{Plx} & \colhead{$\mu_{\rm RA}$} & \colhead{$\mu_{\rm Dec}$} &
	\colhead{Gaia DR2 Source ID} \\
	\colhead{ } & \colhead{(h:m:s)} & \colhead{(d:m:s)} & 
	\colhead{(mag)} & \colhead{(K)} & \colhead{(cm s$^{-2}$)} & 
	\colhead{ } & \colhead{ } & \colhead{ } & \colhead{ } &
	\colhead{(\msun)} & \colhead{(mag)} & \colhead{(kpc)} &
	\colhead{(mas)} & \colhead{(\mas)} & \colhead{(\mas)} &
	\colhead{ }
		}
	\startdata 
J0027$-$1516 &  0:27:51.748 & $-$15:16:26.57 & $17.131 \pm 0.025$ & $10801 \pm 200$ & $6.127 \pm 0.052$ & 1 & 1 & 1 &       1 & $0.176 \pm 0.010$ & $ 8.56 \pm 0.13$  & $0.518 \pm 0.071$ & $ 1.7115 \pm 0.1444$ & $-11.936 \pm 0.2882$ & $-10.126 \pm 0.1991$ & 2374553930375154944 \\
J0042$+$3103 &  0:42:07.253 &    31:03:29.45 & $18.005 \pm 0.016$ & $ 9507 \pm 100$ & $6.274 \pm 0.048$ & 1 & 1 & 1 &       1 & $0.176 \pm 0.010$ & $ 9.32 \pm 0.11$  & $0.545 \pm 0.062$ & $ 1.2825 \pm 0.2221$ & $-14.913 \pm 0.3631$ & $ 2.2823 \pm 0.2752$ &  360595902165353472 \\
J0050$+$2147 &  0:50:46.851 &    21:47:25.66 & $20.061 \pm 0.024$ & $14218 \pm 250$ & $5.826 \pm 0.053$ & 1 & 1 & 0 &       0 & $0.186 \pm 0.010$ & $ 7.12 \pm 0.15$  & $4.102 \pm 0.623$ & $ 0.8389 \pm 0.8569$ & $ 2.9275 \pm 1.7905$ & $-11.545 \pm 2.0285$ & 2801934821646404480 \\
J0124$+$3908 &  1:24:59.733 &    39:08:04.43 & $18.285 \pm 0.013$ & $29175 \pm 330$ & $7.286 \pm 0.047$ & 1 & 0 & 0 &       1 & $0.407 \pm 0.034$ & $ 8.68 \pm 0.12$  & $0.833 \pm 0.104$ & $ 1.0446 \pm 0.2482$ & $ 4.2722 \pm 0.4607$ & $-3.4473 \pm 0.4473$ &  323571256848983552 \\
J0130$+$5321 &  1:30:58.174 &    53:21:38.37 & $14.288 \pm 0.009$ & $ 9231 \pm 100$ & $6.627 \pm 0.056$ & 1 & 1 & 0 &       1 & $0.191 \pm 0.013$ & $10.26 \pm 0.10$  & $0.085 \pm 0.009$ & $ 6.5549 \pm 0.0515$ & $61.1420 \pm 0.0918$ & $-86.462 \pm 0.0855$ &  407508116250828800 \\
J0147$+$0113 &  1:47:20.465 &     1:13:58.28 & $20.216 \pm 0.022$ & $ 9383 \pm 100$ & $6.947 \pm 0.040$ & 1 & 1 & 0 &       0 & $0.240 \pm 0.012$ & $10.76 \pm 0.09$  & $0.809 \pm 0.075$ & $-0.3146 \pm 0.8756$ & $ 3.1155 \pm 1.4860$ & $-52.577 \pm 1.6687$ & 2511132447278844928 \\
J0151$+$1812 &  1:51:20.679 &    18:12:47.95 & $19.604 \pm 0.027$ & $ 8879 \pm  90$ & $6.328 \pm 0.050$ & 1 & 1 & 1 &       1 & $0.154 \pm 0.011$ & $ 9.90 \pm 0.10$  & $0.933 \pm 0.098$ & $-0.1289 \pm 0.4323$ & $13.9995 \pm 0.9192$ & $-2.7449 \pm 0.7981$ &   92092035925893888 \\
J0212$+$2657 &  2:12:16.043 &    26:57:53.52 & $19.419 \pm 0.031$ & $ 9163 \pm 100$ & $6.518 \pm 0.049$ & 1 & 1 & 1 &       0 & $0.170 \pm 0.012$ & $10.14 \pm 0.11$  & $0.804 \pm 0.089$ & $ 1.4057 \pm 0.4510$ & $-9.4283 \pm 0.8561$ & $-13.408 \pm 0.7127$ &  107127651277641472 \\
J0441$-$0547 &  4:41:32.625 &  $-$5:47:34.95 & $18.310 \pm 0.016$ & $12732 \pm 330$ & $5.045 \pm 0.086$ & 0 & 0 & 0 & \nodata & $0.185 \pm 0.011$ & $ 5.40 \pm 0.26$  & $4.733 \pm 1.252$ & $ 0.3092 \pm 0.2872$ & $ 4.8286 \pm 0.4549$ & $-6.2472 \pm 0.3931$ & 3200233905240195968 \\
J0923$-$1218 &  9:23:50.319 & $-$12:18:24.00 & $16.325 \pm 0.004$ & $19455 \pm 210$ & $7.170 \pm 0.041$ & 1 & 0 & 0 &       1 & $0.344 \pm 0.023$ & $ 9.23 \pm 0.13$  & $0.262 \pm 0.034$ & $ 3.6920 \pm 0.0792$ & $-17.484 \pm 0.1362$ & $12.7521 \pm 0.1185$ & 5738500791959712768 \\
J1021$+$0543 & 10:21:53.117 &     5:43:22.28 & $19.360 \pm 0.017$ & $18314 \pm 220$ & $6.703 \pm 0.054$ & 1 & 1 & 1 &       0 & $0.230 \pm 0.013$ & $ 8.60 \pm 0.12$  & $1.420 \pm 0.178$ & $ 1.3622 \pm 0.6033$ & $-10.721 \pm 1.1203$ & $-11.711 \pm 0.8998$ & 3861429723729285376 \\
J1048$-$0000 & 10:48:26.862 &  $-$0:00:56.81 & $18.261 \pm 0.023$ & $ 8484 \pm  90$ & $5.831 \pm 0.051$ & 1 & 1 & 0 &       1 & $0.169 \pm 0.016$ & $ 9.01 \pm 0.25$  & $0.707 \pm 0.175$ & $ 0.6502 \pm 0.3045$ & $-3.0858 \pm 0.4047$ & $-5.3089 \pm 0.3125$ & 3806330138044722176 \\
J1115$+$0246 & 11:15:27.310 &     2:46:21.86 & $18.835 \pm 0.018$ & $27182 \pm 450$ & $7.439 \pm 0.056$ & 1 & 0 & 0 &       0 & $0.446 \pm 0.010$ & $ 9.06 \pm 0.10$  & $0.899 \pm 0.091$ & $ 1.3282 \pm 0.4507$ & $-13.836 \pm 0.7201$ & $-6.2443 \pm 0.4607$ & 3811751005247652352 \\
J1138$-$0035 & 11:38:40.679 &  $-$0:35:32.17 & $14.090 \pm 0.021$ & $31614 \pm 330$ & $5.627 \pm 0.045$ & 0 & 0 & 0 & \nodata & $0.197 \pm 0.010$ & $ 5.30 \pm 0.12$  & $0.571 \pm 0.072$ & $ 0.8649 \pm 0.0630$ & $-8.4262 \pm 0.1173$ & $-25.372 \pm 0.0699$ & 3794197787442075008 \\
J1401$-$0817 & 14:01:18.801 &  $-$8:17:23.43 & $16.456 \pm 0.017$ & $ 8813 \pm  90$ & $5.731 \pm 0.048$ & 1 & 1 & 1 &       0 & $0.216 \pm 0.042$ & $ 7.73 \pm 0.48$  & $0.555 \pm 0.268$ & $ 0.8736 \pm 0.1002$ & $-5.0012 \pm 0.1759$ & $-79.203 \pm 0.1322$ & 3616216816596857984 \\
J1545$+$4301 & 15:45:21.102 &    43:01:41.85 & $18.998 \pm 0.021$ & $ 9707 \pm 110$ & $6.222 \pm 0.043$ & 1 & 1 & 1 &       1 & $0.174 \pm 0.010$ & $ 9.13 \pm 0.10$  & $0.939 \pm 0.100$ & $ 0.7866 \pm 0.2128$ & $ 0.4223 \pm 0.3776$ & $ 0.5518 \pm 0.4952$ & 1396245695576598272 \\
J1638$+$3500 & 16:38:26.274 &    35:00:12.03 & $14.561 \pm 0.015$ & $37250 \pm 570$ & $8.070 \pm 0.050$ & 1 & 0 & 0 &       1 & $0.698 \pm 0.030$ & $ 9.49 \pm 0.44$  & $0.103 \pm 0.046$ & $ 6.8981 \pm 0.0321$ & $-35.102 \pm 0.0512$ & $ 8.9790 \pm 0.0652$ & 1327577144269234176 \\
J1708$+$2225 & 17:08:16.358 &    22:25:51.07 & $19.106 \pm 0.015$ & $22343 \pm 450$ & $6.865 \pm 0.059$ & 1 & 0 & 0 &       1 & $0.320 \pm 0.011$ & $ 8.29 \pm 0.08$  & $1.612 \pm 0.130$ & $ 1.2867 \pm 0.4295$ & $-1.7690 \pm 0.6684$ & $ 3.5902 \pm 0.6154$ & 4568269229123390336 \\
J1738$+$2927 & 17:38:35.467 &    29:27:50.63 & $19.309 \pm 0.017$ & $12018 \pm 230$ & $6.972 \pm 0.051$ & 1 & 1 & 1 &       1 & $0.261 \pm 0.016$ & $10.01 \pm 0.11$  & $0.780 \pm 0.090$ & $ 1.1932 \pm 0.3759$ & $-9.8331 \pm 0.5359$ & $ 6.1181 \pm 0.7191$ & 4595849099618519680 \\
J2147$+$1859 & 21:47:28.476 &    18:59:59.76 & $19.580 \pm 0.022$ & $ 9618 \pm 110$ & $5.639 \pm 0.059$ & 1 & 1 & 1 &       1 & $0.157 \pm 0.021$ & $ 7.87 \pm 0.13$  & $2.199 \pm 0.286$ & $ 0.8328 \pm 0.6796$ & $-0.7578 \pm 0.9284$ & $-2.6692 \pm 1.1753$ & 1780334519094674304 \\
J2245$+$0750 & 22:45:21.283 &     7:50:48.74 & $19.635 \pm 0.022$ & $10782 \pm 110$ & $6.184 \pm 0.056$ & 1 & 1 & 1 &       1 & $0.178 \pm 0.010$ & $ 8.69 \pm 0.14$  & $1.547 \pm 0.213$ & $ 0.1356 \pm 0.7483$ & $ 9.0727 \pm 1.2676$ & $-3.2587 \pm 1.0968$ & 2712813082023657600 \\
J2317$+$0602 & 23:17:57.418 &     6:02:52.09 & $19.494 \pm 0.035$ & $12043 \pm 160$ & $7.441 \pm 0.052$ & 1 & 0 & 0 &       1 & $0.381 \pm 0.029$ & $10.76 \pm 0.09$  & $0.558 \pm 0.054$ & $ 2.1029 \pm 0.6595$ & $ 7.1097 \pm 2.4904$ & $-2.1619 \pm 0.9112$ & 2664126329188074240 \\
J2332$+$0427 & 23:32:46.564 &     4:27:35.20 & $18.022 \pm 0.014$ & $11967 \pm 160$ & $5.834 \pm 0.048$ & 1 & 1 & 1 &       0 & $0.181 \pm 0.010$ & $ 8.06 \pm 0.18$  & $1.087 \pm 0.199$ & $ 0.6446 \pm 0.2393$ & $14.4921 \pm 0.4318$ & $-13.927 \pm 0.2615$ & 2660056212019666688 \\
J2339$+$2024 & 23:39:53.667 &    20:24:44.84 & $18.244 \pm 0.014$ & $ 8019 \pm  90$ & $5.263 \pm 0.059$ & 0 & 0 & 0 & \nodata & $0.182 \pm 0.013$ & $ 7.53 \pm 0.08$  & $1.387 \pm 0.107$ & $ 0.1180 \pm 0.2031$ & $-0.3283 \pm 0.3558$ & $-2.3290 \pm 0.2247$ & 2826170531823332096 \\
J2339$-$0347 & 23:39:38.450 &  $-$3:47:34.51 & $18.542 \pm 0.025$ & $16047 \pm 260$ & $5.982 \pm 0.047$ & 1 & 1 & 1 &       1 & $0.188 \pm 0.016$ & $ 7.28 \pm 0.12$  & $1.882 \pm 0.223$ & $ 0.1387 \pm 0.6493$ & $-5.3204 \pm 1.1572$ & $-3.1589 \pm 0.7429$ & 2639275992010565376 \\
	\enddata
	\tablecomments{$g_0$ is de-reddened SDSS $g$-band apparent magnitude, except 
for 5 cases when it is derived from PanStarrs $g$ or {\it Gaia} $G$.  Measured 
\teff\ and \logg\ values are corrected for 3D effects following \citet{tremblay15}.
	Classifications are set to 1 if true or 0 if 
false, i.e.\ WD=1 indicates a WD, ELM=1 indicates an ELM WD, Clean=1 indicates an 
ELM WD in the clean sample, and disk=1 indicates an object that orbits in the disk.  
	WD mass, absolute $g$-band magnitude $M_g$, and distance are derived using 
the models of \citet{althaus13} and \citet{istrate16}, but are only valid for 
objects marked WD=1.
	For the eight candidates without {\it Gaia} 5-parameter solutions, we list 
proper motions from {\it Gaia}-PanStarrs1 \citep{tian17}.
	This table is available in its entirety in machine-readable and Virtual 
Observatory forms in the online journal. The 25 well-constrained new binaries are 
shown here for guidance regarding form and content. }
 \end{splitdeluxetable*}

\subsection{White Dwarf Parameters}

	For every candidate, we interpolate its \teff\ and \logg\ measurements 
through WD evolutionary tracks to estimate its putative WD mass and luminosity.
	We use \citet{istrate16} tracks for ELM WDs because they are computed for 
both solar metallicity and halo metallicity progenitors.  A significant fraction of 
the observed ELM WDs belong to the halo.  We also use \citet{althaus13} tracks, and 
in one case \citet{tremblay11} tracks, to cover the full range of temperature and 
surface gravity of our sample.

	The \citet{istrate16} tracks overlap the observations in the region 
8,$800~{\rm K} < T_{\rm eff} < 22,$000~K and $\log{g}<7.1$ (the cyan box in 
Figure~\ref{fig:teff}).  {\it This motivates us to refer to these candidates as ELM 
WDs.} In this region, we apply \citet{istrate16} $Z=0.02$ tracks with rotation to 
disk objects and the $Z=0.001$ tracks with rotation to halo objects.  We apply 
\citet{althaus13} tracks to everything else, except for the more massive WD 
J1638+3500 which requires \citet{tremblay11} tracks.  The WD masses derived from the 
\citet{istrate16} and \citet{althaus13} tracks differ by 0.00$\pm$0.012 \msun\ in 
their region of overlap.  We thus compute mass errors by propagating the \teff\ and 
\logg\ uncertainties through the tracks and adding $\pm$0.01 \msun\ in quadrature.

	We then compute heliocentric distances, $d$, using de-reddened apparent SDSS 
$g$-band magnitude, $g_0$, and the absolute magnitude $M_g$ derived from the tracks, 
$d=10^{((g_0-M_g)/5-2)}$ kpc.  Applying the full reddening correction may be 
incorrect for the nearest WDs; however the median WD in our sample is 0.8~kpc 
distant and has low $E(B-V)=0.031$ mag reddening.

	Figure~\ref{fig:plxplx} compares {\it Gaia} parallax to the inverse of our 
spectrophotometric distance estimate.  We see two bands in Figure~\ref{fig:plxplx}.
	The band of candidates near the 1:30 ratio line in Figure~\ref{fig:plxplx} 
are approximately 30 times more distant, or $\sim$1,000 times more luminous, than we 
estimate from WD models.  Since \teff\ should be accurately measured, we conclude 
that these candidates have radii $\sim$30 times larger than WDs.  The candidates 
near the 1:30 ratio line are thus likely metal-poor stars at kpc distances in the 
halo, objects traditionally called field blue stragglers at these temperatures.

	The candidates scattered around the diagonal 1:1 ratio line in 
Figure~\ref{fig:plxplx} are likely WDs.  There are also some candidates just below 
the 1:1 ratio line which notably have $T_{\rm eff}<9,$000~K.  If we again assume 
temperature is robust, these cool WDs just below the 1:1 ratio line can be explained 
by either $\sim$1.7$\times$ inflated radii or $\sim$0.5 dex systematic \logg\ 
errors.  The latter explanation is consistent with our previously published 
systematic error estimate for pure hydrogen models at $<$9,000~K temperatures 
\citep{brown17a}.

	For the candidates with $T_{\rm eff}>9,$500~K, the ELM WD models of 
\citet{istrate16} and \citet{althaus13} provide remarkably accurate measures of 
luminosity.  The mean parallax ratio of the 35 candidates with $\pi/\sigma_\pi>$5 
and 9,$500<T_{\rm eff}<30,$000~K, after clipping a single field blue straggler
interloper, is $0.97\pm0.04$.

\subsection{Clean ELM WD Sample}

	We define a clean sample of WDs, and of ELM WDs, on the basis of our stellar 
atmosphere measurements and {\it Gaia} parallax.  The subset of our sample with 
$\pi/\sigma_\pi>$5 demonstrate that $T_{\rm eff} > 9,$000~K and $\log{g}>5.5$ 
candidates are a clean set of WDs.  Metal-poor main sequence stars at the same 
temperatures have distinct $\log{g}\le4.7$ \citep[e.g.,][]{marigo17}.

	Thus we start building our clean sample of WDs from the 115 candidates with 
$T_{\rm eff} > 9,$000~K and $\log{g}>5.5$.  We remove 5 objects that do not belong: 
the sdB star, and four candidates with $\pi/\sigma_\pi>$5 and distance estimates 
that differ by more than 3$\times$.  

	We then add candidate WDs with $T_{\rm eff} < 9,$000~K or $\log{g}< 5.5$ on 
the basis of parallax and binary orbital period.  Excluding the sdA pulsator 
J1355+1956 \citep{bell17}, six candidates have $\pi/\sigma_\pi>$5 and distance
estimates that agree to within a factor of 3.  Interestingly, two-thirds are 
short-period binaries.  There are an additional six candidates with significant 
$k>100$~\kms\ orbital motion and short $P<0.27$~d periods.  Orbits with $P<0.27$~d 
exclude metal-poor A-type stellar radii on basis of Padova tracks \citep{marigo17} 
and the Roche lobe criterion \citep{eggleton83}.  Summed together, the result is a 
sample of 122 likely WDs.  We label these objects WD=1 in Table~\ref{tab:param}.

	We identify ELM WDs as those WDs with $M<0.3$ \msun, in other words, the WDs 
that overlap the ELM WD tracks \citep{althaus13, istrate16}.  There are 79 ELM WDs 
in our sample by this definition.  Two WDs with masses just above 0.3 \msun, 
J0822+3048 and J0935+4411, are now excluded.  However, some of the ELM WDs included 
in our definition are drawn from outside the SDSS survey footprint, or have apparent 
magnitudes outside our primary magnitude selection.

	Thus we additionally define a clean ELM WD sample:  ELM WDs in the 
de-reddened magnitude range $15<g_0<20$, located in the SDSS footprint, with 
8,$800~{\rm K} < T_{\rm eff} < 22,$000~K and $5.5\le\log{g}\le7.1$.  We choose this 
range to maximize the overlap between the ELM WD tracks and the observations, and to 
minimize contamination (see Figure~\ref{fig:teff}).  This excludes ancillary 
candidates we identified in the TESS Input Catalog, the Edinburgh-Cape Survey, and 
the LAMOST catalog so that the photometric and spatial selection is uniform. We use 
inclusive \logg\ boundaries to include the eclipsing ELM WD binary J0751$-$0141 
\citep{kilic14b}.  The clean sample of ELM WDs contains 62 objects and is 
essentially complete within our selection criteria.

	Table~\ref{tab:param} summarizes our classifications.  The values of each 
column are set to 1 if true or 0 if false, i.e.\ ELM=1 indicates an ELM WD, and 
Clean=1 indicates an ELM WD in the clean sample.  Note that the clean ELM WD sample 
defined here differs from our previous papers: we intentionally exclude the coolest 
and lowest-gravity ELM WD candidates so as to minimize contamination from other 
stellar populations.

 \begin{figure*} 
 \plottwo{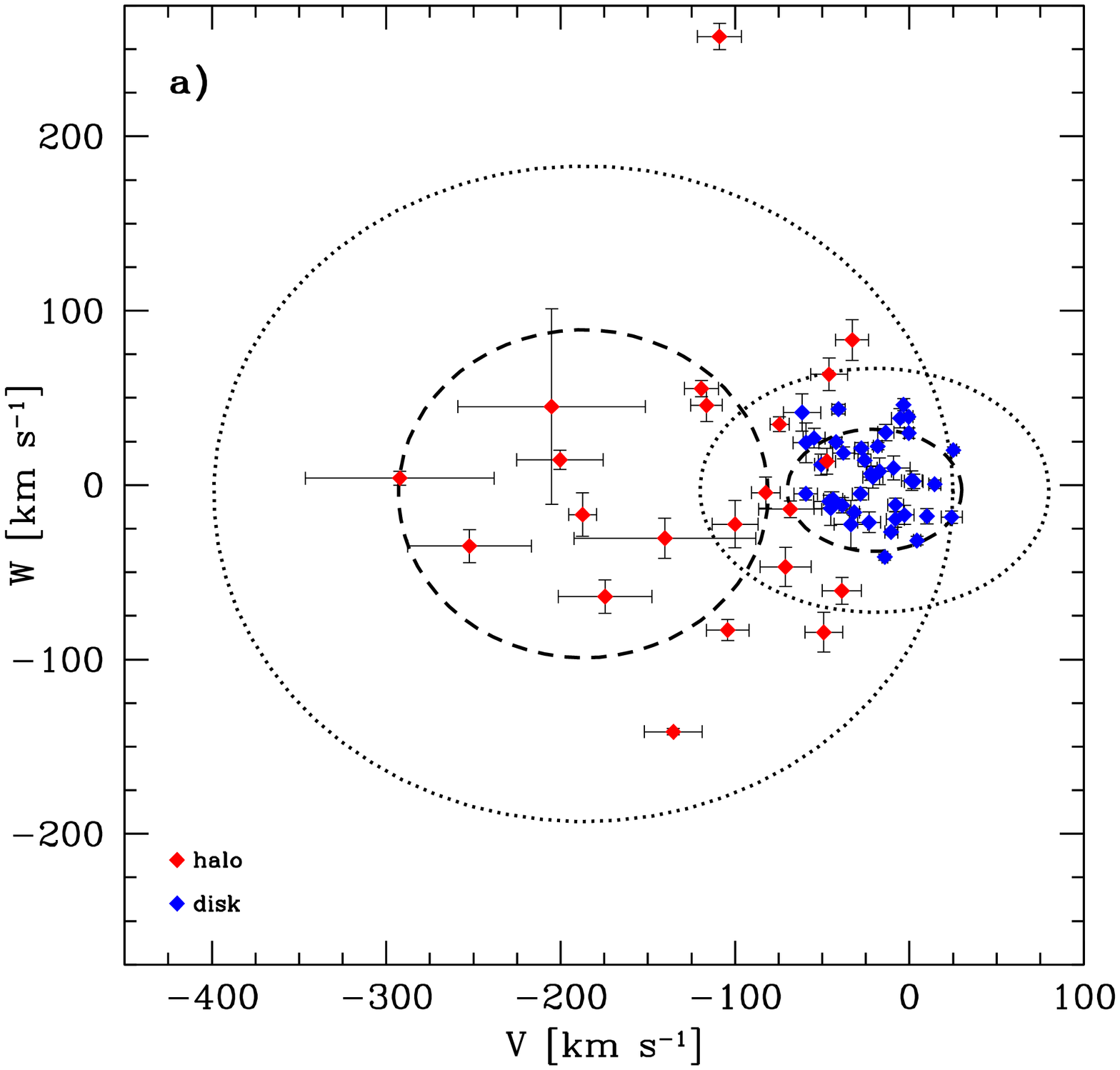}{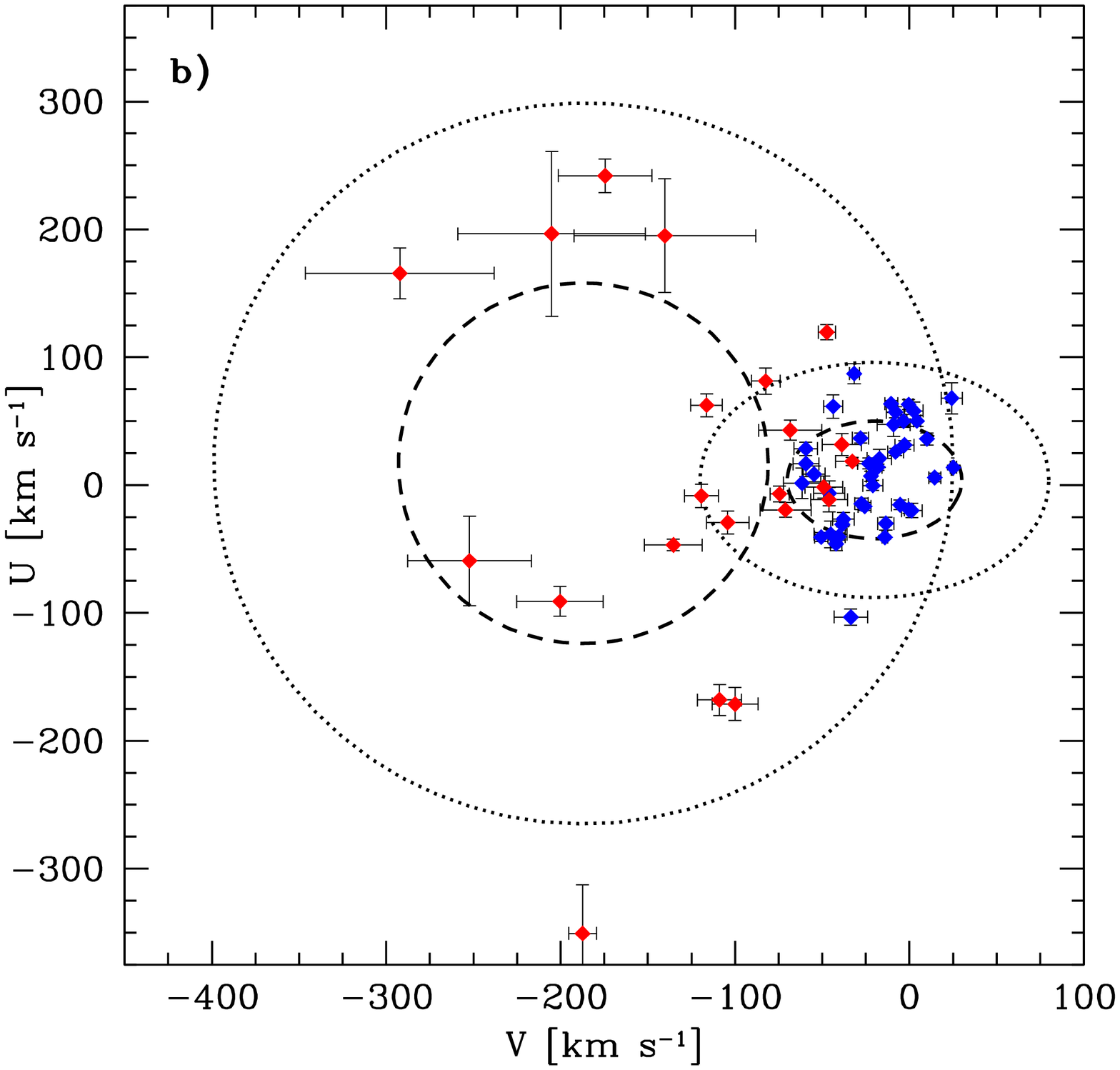} 
 \caption{ \label{fig:uvw}
	Velocity distribution of the clean ELM WD sample, plotted in Galactic 
cartesian velocity components $U$ (in the direction of the Galactic center), $V$ (in 
the direction of Galactic rotation), and $W$ (in the direction of the north Galactic 
pole).  For comparison are the 1$\sigma$ (dashed) and 2$\sigma$ (dotted) velocity 
ellipsoid values for stellar thick disk and halo populations \citep{chiba00}.  We 
classify 37\% of the sample as halo (red points) and 63\% as disk (blue points). }
 \end{figure*}

\subsection{White Dwarf Disk/Halo Membership}

	We classify the disk/halo membership for the clean ELM WD sample on the 
basis of space velocity.  Previously, we found that 37\% of ELM WD binaries in our 
sample orbit in the halo \citep{gianninas15, brown16b}.  {\it Gaia} proper motions 
provide an order-of-magnitude improvement in accuracy compared to previous work.

	We compute tangential velocity from the product of {\it Gaia} proper motion
and spectrophotometric distance, because spectrophotometric distance has smaller
uncertainties than parallax for most of the clean ELM WD sample.  We measure
systemic radial velocity directly, and correct it for ELM WD gravitational redshift.  
The median tangential and systemic radial velocity errors in the clean ELM WD sample
are 20~\kms\ and 4~\kms, respectively.

	We calculate Galactic $UVW$ velocities assuming a circular motion of 
235~\kms\ \citep{reid09} and the solar motion of \citet{schonrich10}, and determine 
disk/halo membership on the basis of ELM WD space velocity and spatial location 
using equations 2--8 in \citet{brown16b}.  This approach yields 35\% (22/62) halo 
objects and 65\% (40/62) disk objects in the clean ELM WD sample, essentially the 
same fraction as before.  We present the disk/halo classifications for all 122
WDs in Table~\ref{tab:param}.

	Figure~\ref{fig:uvw} plots the distribution of Galactic $U$, $V$, and $W$ 
velocity components for the clean ELM WD sample.  Disk objects are drawn in blue, 
halo objects are drawn in red.  For comparison, we draw the velocity ellipsoid 
values of the halo and thick disk from \citet{chiba00}.

	Interestingly, disk and halo ELM WDs exhibit statistically identical
distributions of other parameters.  Disk and halo ELM WDs overlap in \teff\ and
\logg\ space.  A two-sample Anderson-Darling test \citep{scholz87} on the ELM WD
mass distribution, semi-amplitude distribution, and orbital period distribution all
have p-values around 0.4.  We conclude that ELM WDs share similar binary properties, 
described below, regardless of their disk/halo origin.

\section{BINARIES}

	We now focus on the binaries.  We present orbital parameters for 128 
binaries in the completed ELM Survey, including 29 published here for the first time 
(25 of which are well-constrained).  We use follow-up radial velocity measurements 
to rule out period aliases in previously published binaries, and use X-ray and radio 
observations to place constraints on the presence of millisecond pulsar companions 
around previously published binaries.  We close with an optical light curve for the 
new $P=0.048$~d ELM WD binary J1738+2927.

\begin{figure*} 
 \plotone{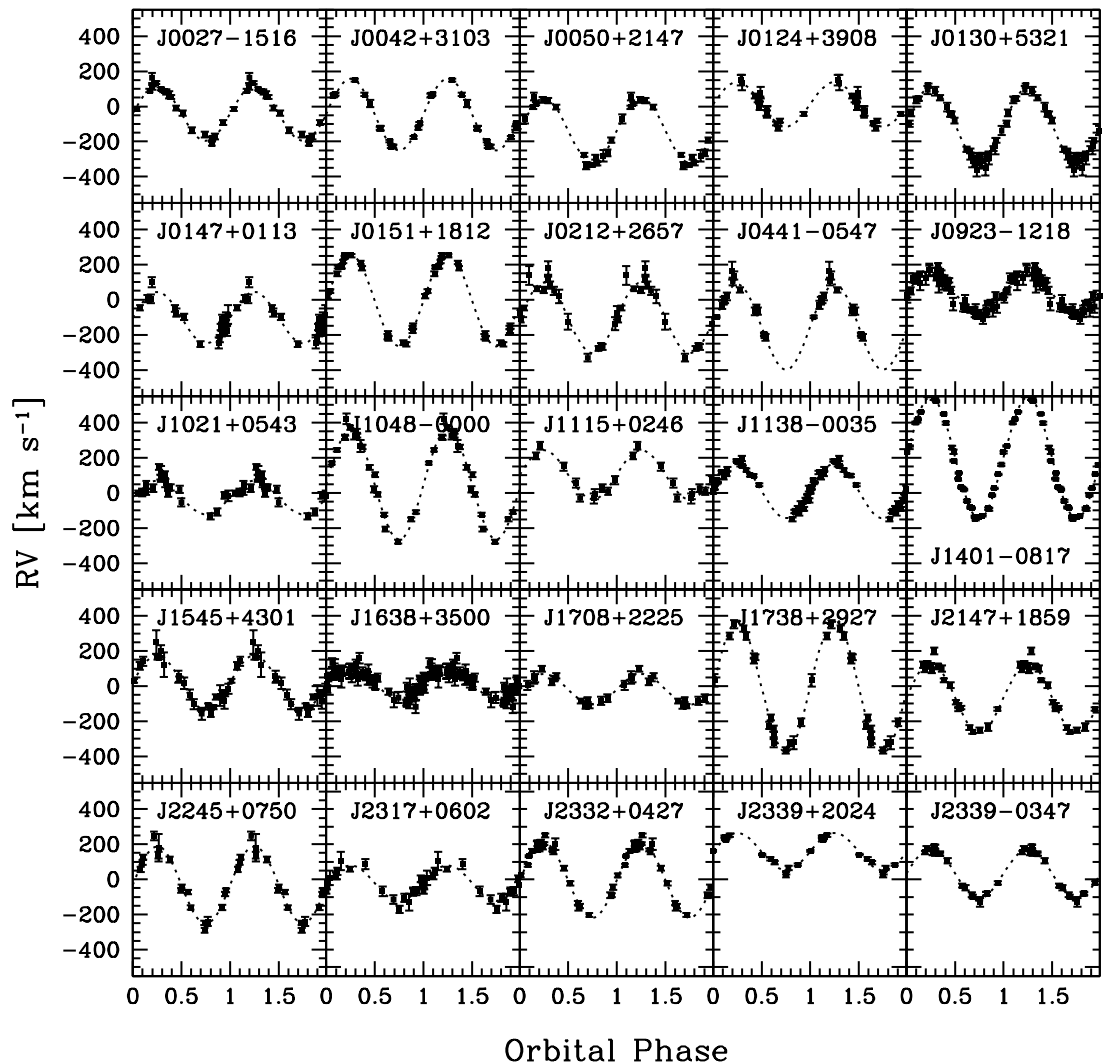} 
 \caption{ \label{fig:orb}
	Radial velocities phased to the best-fit orbital solutions for the 25 
well-constrained new binaries. } 
 \end{figure*}

\subsection{Velocity Variability}

	We identify binaries among the low mass WD candidates on the basis of 
radial velocity variability.  Radial velocities are measured with the 
cross-correlation technique using the full wavelength range of the spectra, as 
described in Section 2.3.  A pair of radial velocities is often sufficient to detect 
the median $P=6$~h, $k=200$~\kms\ binary in our sample.  However, we find that 4 to 
7 observations are necessary to perform a significant test for orbital motion.  We 
use the F-test to quantify whether the variance of the observations, given 
measurement errors, is consistent with a constant velocity.  Candidates with 
p-values $<0.01$ are inconsistent with constant velocity, in other words, they are 
likely binaries.

	Sensitivity tests show that our cadence of observations and measurement 
errors have a 99\% likelihood of detecting binaries with semi-amplitudes 
$k>100$~\kms\ and $P<2$~d \citep{brown16a}.  We acquire a median 21 observations per 
binary.  Observations are separated by minutes to hours over the course of multiple 
observing runs; the exact cadence of observations depends on the period of the 
binary and where it was placed on the sky during our observing runs.

	We find that 128 of the 230 candidates have statistically significant 
velocity variability.  These binaries are drawn with green symbols in Figures 
\ref{fig:teff}--\ref{fig:plxplx}.  Of the 128 velocity variable objects, 99 are 
published in our previous papers and 29 are new (25 of which are well-constrained 
binaries).

\begin{deluxetable*}{lcccccccccc}  	
 \tabletypesize{\scriptsize} \tablecolumns{11} \tablewidth{0pt} 
 \tablecaption{Binary Parameters\label{tab:orb}} 
 \tablehead{
   \colhead{Object} & \colhead{$N_{\rm obs}$} & \colhead{$P$} & \colhead{$k$} 
   & \colhead{$\gamma$} & \colhead{$M_{\rm 2,min}$} & \colhead{$\log{\tau_{\rm ,max}}$} 
   & \colhead{alias?} & \colhead{$\Delta\chi^2_{\rm alias}$} 
   & \colhead{$P_{\rm alias}$} & \colhead{$h_c\sqrt{(4~{\rm yr})f}$} \\ 
   & & (d) & (km s$^{-1}$) & (km s$^{-1}$) & (\msun) & (yr) & & & (days) 
   & ($\times10^{-21}$) }
	\startdata 
J0027$-$1516 & 18 & $0.42458 \pm 0.00014$ & $155.4 \pm  6.3$ & $ -42.4 \pm  6.9$ & 0.36 & 10.79 & 0 & 64.17 & 0.29780 & $ 1.4^{+ 2.0}_{- 0.6}$ \\
J0042$+$3103 & 16 & $0.29725 \pm 0.00018$ & $204.2 \pm  5.2$ & $ -48.6 \pm  5.5$ & 0.49 & 10.28 & 0 & 338.9 & 0.22913 & $ 2.4^{+ 3.3}_{- 0.9}$ \\
J0050$+$2147 & 15 & $0.36059 \pm 0.00002$ & $183.7 \pm  6.6$ & $-138.9 \pm 10.9$ & 0.46 & 10.51 & 0 & 55.84 & 0.05355 & $0.26^{+0.36}_{-0.10}$ \\
J0124$+$3908 & 16 & $1.29211 \pm 0.00433$ & $127.0 \pm  9.9$ & $   1.8 \pm 11.3$ & 0.69 & 11.54 & 1 &  0.74 & 0.22477 & $0.77^{+0.92}_{-0.30}$ \\
J0130$+$5321 & 32 & $0.19205 \pm 0.00020$ & $209.1 \pm  5.1$ & $-105.8 \pm  8.8$ & 0.40 &  9.81 & 0 & 433.6 & 0.23789 & $  24^{+  35}_{-  10}$ \\
J0147$+$0113 & 19 & $1.30338 \pm 0.00483$ & $145.9 \pm 15.7$ & $-107.9 \pm 15.2$ & 0.74 & 11.74 & 1 &  6.95 & 0.57599 & $0.50^{+0.57}_{-0.19}$ \\
J0151$+$1812 & 19 & $0.14812 \pm 0.00001$ & $259.8 \pm  3.5$ & $  -8.1 \pm  3.1$ & 0.47 &  9.54 & 0 & 154.3 & 0.13020 & $ 2.8^{+ 3.8}_{- 1.1}$ \\
J0212$+$2657 & 17 & $0.44908 \pm 0.00197$ & $202.0 \pm 11.5$ & $-107.1 \pm  9.3$ & 0.62 & 10.70 & 0 & 78.62 & 0.31291 & $ 1.1^{+ 1.4}_{- 0.4}$ \\
J0441$-$0547 & 15 & $1.31997 \pm 0.00060$ & $242.7 \pm 18.1$ & $-155.9 \pm 11.7$ & 2.28 & 11.51 & 1 &  6.54 & 1.55179 &         \nodata        \\
J0923$-$1218 & 51 & $0.14896 \pm 0.00002$ & $117.0 \pm  3.7$ & $  29.4 \pm  2.7$ & 0.19 &  9.56 & 0 & 64.73 & 0.17512 & $  31^{+  28}_{-  10}$ \\
J1021$+$0543 & 20 & $1.24995 \pm 0.00410$ & $ 95.6 \pm 11.6$ & $ -30.0 \pm 13.6$ & 0.33 & 11.98 & 0 & 14.99 & 1.38071 & $0.17^{+0.25}_{-0.07}$ \\
J1048$-$0000 & 20 & $0.12063 \pm 0.00001$ & $312.8 \pm  8.1$ & $  45.7 \pm  6.1$ & 0.62 &  9.18 & 0 & 751.1 & 0.10763 & $ 6.3^{+ 7.7}_{- 2.6}$ \\
J1115$+$0246 & 10 & $0.12405 \pm 0.00004$ & $139.9 \pm 12.2$ & $  90.0 \pm  8.1$ & 0.26 &  9.15 & 1 &  0.39 & 0.14175 & $  16^{+  14}_{-   5}$ \\
J1138$-$0035 & 36 & $0.20769 \pm 0.00002$ & $155.0 \pm  4.9$ & $   9.9 \pm  3.9$ & 0.25 & 10.05 & 0 & 74.25 & 0.17189 &         \nodata        \\
J1401$-$0817 & 35 & $0.11299 \pm 0.00001$ & $346.2 \pm  2.7$ & $ 198.7 \pm  2.0$ & 0.79 &  8.93 & 0 &  6885 & 0.12746 & $  13^{+  19}_{-   7}$ \\
J1545$+$4301 & 25 & $0.30931 \pm 0.00016$ & $154.8 \pm  4.1$ & $  18.2 \pm  4.0$ & 0.30 & 10.50 & 0 & 111.0 & 0.45533 & $0.94^{+1.45}_{-0.37}$ \\
J1638$+$3500 & 66 & $0.90606 \pm 0.00031$ & $ 89.5 \pm  4.4$ & $ -17.5 \pm  4.1$ & 0.45 & 11.09 & 0 & 24.60 & 0.47468 & $  30^{+  30}_{-  12}$ \\
J1708$+$2225 & 17 & $0.23735 \pm 0.00024$ & $115.5 \pm  8.5$ & $  -6.5 \pm  6.6$ & 0.22 & 10.07 & 1 &  8.56 & 1.00795 & $ 2.3^{+ 2.3}_{- 0.8}$ \\
J1738$+$2927 & 17 & $0.04770 \pm 0.00011$ & $372.7 \pm 13.2$ & $ -11.9 \pm 12.8$ & 0.55 &  7.97 & 0 & 66.08 & 0.05274 & $  24^{+  29}_{-   9}$ \\
J2147$+$1859 & 20 & $0.12879 \pm 0.00002$ & $198.3 \pm  6.6$ & $ -67.9 \pm  5.5$ & 0.27 &  9.56 & 0 & 153.0 & 0.17977 & $0.94^{+1.48}_{-0.40}$ \\
J2245$+$0750 & 18 & $0.39664 \pm 0.00102$ & $220.5 \pm 10.1$ & $ -34.2 \pm 11.7$ & 0.70 & 10.50 & 0 & 73.41 & 0.65750 & $0.77^{+0.89}_{-0.28}$ \\
J2317$+$0602 & 20 & $0.86702 \pm 0.00133$ & $100.7 \pm  7.3$ & $ -34.2 \pm 12.2$ & 0.38 & 11.32 & 1 &  5.47 & 1.27191 & $ 1.4^{+ 1.8}_{- 0.6}$ \\
J2332$+$0427 & 24 & $0.36792 \pm 0.00009$ & $212.5 \pm  4.9$ & $  -7.0 \pm  3.5$ & 0.61 & 10.44 & 0 & 420.6 & 0.51537 & $ 1.1^{+ 1.4}_{- 0.4}$ \\
J2339$+$2024 & 14 & $0.79578 \pm 0.00008$ & $106.3 \pm  5.0$ & $ 157.3 \pm  3.7$ & 0.28 & 11.60 & 0 & 56.46 & 0.43107 &         \nodata        \\
J2339$-$0347 & 19 & $0.67069 \pm 0.00078$ & $139.7 \pm  6.0$ & $  31.2 \pm 10.9$ & 0.41 & 11.26 & 0 & 73.22 & 1.37843 & $0.26^{+0.37}_{-0.10}$ 
	\enddata
	\tablecomments{$N_{\rm obs}$ is the number of spectroscopic observations. 
$P$ is the binary orbital period. $k$ is the radial velocity semi-amplitude. 
$\gamma$ is the systemic velocity corrected for gravitational redshift. $M_{\rm 
2,min}$ is the minimum mass of the secondary. $\tau_{\rm ,max}$ is the maximum 
gravitational merger timescale.  Binaries with significant period aliases have 
alias=1 if true or 0 if false. $\Delta\chi^2_{\rm alias}$ is the $\chi^2$ at $P_{\rm 
alias}$ relative to the global $\chi^2$ minimum.  $h_c$ is the characteristic strain 
and $\sqrt{(4~{\rm yr})f}$ is the S/N boost from the number of cycles during the 
{\it LISA} observation time (the values plotted in Figure~\ref{fig:gw}).  This table 
is available in its entirety in machine-readable and Virtual Observatory forms in 
the online journal.  The 25 well-constrained new binaries are shown here for 
guidance regarding form and content. }
  \end{deluxetable*}

\subsection{Binary Orbital Elements}

	We calculate orbital elements as described in previous ELM Survey papers.  
We start by using the summed, rest-frame spectrum of each target as its own 
cross-correlation template to maximize the velocity precision.  We then minimize 
$\chi^2$ for a circular orbit fit following the code of \citet{kenyon86}.  We find 
that our cross-correlation approach underestimates velocity error, however.  To 
obtain a reduced $\chi^2\simeq1$ requires adding a median 15~\kms\ velocity error in 
quadrature to the measurements.  Thus we estimate orbital element errors by 
re-sampling the velocities with the extra error added in quadrature and re-fitting 
the orbital solution 10,000 times.  This Monte Carlo approach samples the $\chi^2$ 
space in a self-consistent way.  We report orbital element errors derived from the 
15.9\% and 84.1\% percentiles of the distributions in Table~\ref{tab:orb}.  
Systemic velocities are corrected for gravitational redshift using the WD parameters 
in Table~\ref{tab:param}.

	Figure~\ref{fig:orb} plots the radial velocities of the 25 well-constrained 
new binaries, phased to their best-fit orbits.  Four other objects have significant 
velocity variability but are not WDs on the basis of their parallax, so we did 
not pursue a full set of observations.

	The 25 well-constrained new binaries with robust orbital solutions are 
mostly low mass WD binaries, including two previously unknown binaries that we 
selected as additional low mass WD candidates:  we found J0923$-$1218 = 
WD~0921$-$120 in the Edinburgh-Cape Survey \citep{kilkenny97}, and J0130+5321 
= GD~278 in the {\it TESS} bright WD catalog \citep{raddi17}.  Two objects are 
not low mass WD binaries but we publish the observations for completeness:  
J1638+3500 is a hot 0.7 \msun\ WD we observed for the SWARMS survey 
\citep{badenes09}, and J1138$-$0035 turns out to be a hot subdwarf B star 
\citep{geier11}.  In 6 cases the orbital solutions have period aliases due to 
insufficient sampling (as seen in J1115+0246 and J1708+2225) and/or uneven phase 
coverage (as seen in J0124+3908 and J0441$-$0547).

	Period aliases, not statistical errors, are the largest source of 
uncertainty in the orbital solutions.  We consider an object to have a significant 
period alias if its orbital elements have local $\chi^2$ minima within $\Delta 
\chi^2 = 13.3$ of the global $\chi^2$ minimum \citep{press92}.  On this basis, 27\% 
(34/128) of the binaries have significant period aliases.  Many of the binaries with 
period aliases are field blue stragglers that we chose not to continue observing; 
only 15\% (15/98) of the WD binaries have period aliases.  

	For completeness, we present the strongest period alias, and its $\Delta 
\chi^2$ value with respect to the global minimum, for all 128 binaries in 
Table~\ref{tab:orb}.  The aliases are found equally at longer and shorter periods.  
The exceptions are low semi-amplitude field blue stragglers.  These objects 
often have short $\sim$1~h period aliases trivially matched to the cadence of 
observations, but their low semi-amplitudes suggest the long period solution is 
likely correct.

	We obtained additional observations that eliminated period aliases for 11 
previously published WD binaries.  In seven cases, the originally published period 
was correct and the orbital solution is unchanged.  In four cases, the alias was 
correct:  J1005+0542 is $P=4.5$~h, J1422+4352 is $P=14.9$~h, J1439+1002 is 
$P=18.6$~h, and J1557+2823 is $P=6.9$~h.

\begin{figure} 
 \includegraphics[width=3.5in]{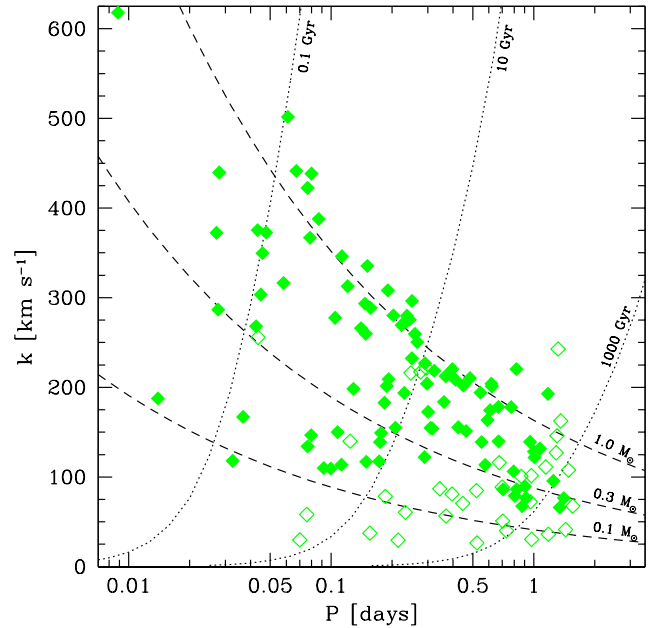} 
 \caption{ \label{fig:pk}
	Observed semi-amplitude and orbital period of the 128 binaries.  Binaries 
with period aliases are drawn with a single open symbol at their best-fit period.  
Statistical errors are smaller than the symbol size.  For the purpose of 
guidance, dashed lines indicate the approximate companion mass and dotted lines 
indicate the corresponding gravitational wave merger timescale, calculated assuming 
the visible star is a 0.2~\msun\ WD as described in the text.
 } \end{figure}

	Figure~\ref{fig:pk} plots the overall distribution of velocity 
semi-amplitude, $k$, versus orbital period, $P$, for all 128 binaries.  Binaries 
with period aliases are drawn with a single open symbol at their best-fit period.  
For the purpose of guidance, not analysis, we draw dashed lines that indicate
the approximate companion mass calculated for $M_1=0.2$ \msun\ and inclination 
$i=60\arcdeg$. The lines in Figure~\ref{fig:pk} are thus only relevant to the 
ELM WD systems.  Dotted lines indicate the approximate gravitational wave merger 
timescale calculated with the same assumptions.  Most of the binaries with period 
aliases are field blue stragglers that have $k<75$~\kms\ or $P>30$~h.  The outlier 
at $(P,k)=(1.3$~d, 243 \kms), J0441$-$0547, has a significant period alias at 
0.57~d.

	In the absence of a constraint on inclination, i.e.\ from eclipses, our 
radial velocity measurements determine only the minimum mass of the companion, 
$M_2$.  We list minimum $M_2$ mass values for the best-fit periods in 
Table~\ref{tab:orb}.  The unseen companions in the 98 binaries containing a 
visible WD all have minimum $M_2$ masses consistent with being other WDs.  
Spectral energy distributions provide no further constraint because, by design, we 
observe candidates dominated by the light of a low mass WD.  The $0.97\pm0.04$ ratio 
of {\it Gaia} parallax to inverse spectro-photometric distance, reported above, 
affirms that the companions are significantly fainter than the visible low mass WD.
The low mass WDs with the largest minimum $M_2$ values and no period aliases 
are J0802$-$0955 and J0811+0225, which have $M_2>1.2$~\msun.

	The minimum companion mass allows us to calculate the maximum gravitational 
wave merger timescale, $\tau$, of the binaries.  We list the maximum merger 
timescales in Table~\ref{tab:orb}.  The values of $\tau$ range from $10^6$~yr to 
$10^{12}$~yr for our sample of binaries, as illustrated in Figure~\ref{fig:pk}.

\subsection{Millisecond Pulsar Companions} 

	Given the unknown inclination of our single-lined spectroscopic binaries, 
some of the unseen companions may be neutron stars.  If true, binary evolution 
should naturally produce millisecond pulsars.  Indeed, millisecond radio pulsars are 
commonly observed with low mass WD companions \citep{manchester05, vankerkwijk05}.  
Low mass WD+pulsar binaries have orbital periods of hours to days similar to the 
binaries observed here \citep{vankerkwijk96, antoniadis13}.

	Millisecond pulsars have wide radio beams that cover $\sim$80\% of the sky 
\citep{lyne88} and invariably show thermal X-ray emission from the heated neutron 
star polar caps; one pole should be visible from any observing angle due to 
gravitational bending of light rays \citep{beloborodov02}.  Because millisecond 
pulsars have lifetimes exceeding $10^{10}$ years, any putative millisecond pulsar 
companion should be active now.

	These facts motivated our follow-up radio and X-ray search for millisecond 
pulsar companions in the ELM Survey.  Previous searches for millisecond pulsar 
companions to known low mass WDs, mostly in the ELM Survey, have yielded no neutron 
star counterparts \citep{vanleeuwen07, agueros09a, agueros09b, kilic13, kilic14b, 
kilic16, andrews18}.  We present our final set of radio and X-ray observations 
of low mass WDs in the ELM Survey sample.

\subsubsection{Radio Observations}

	We engaged in a long-term radio campaign using the Robert C.\ Byrd Green 
Bank Telescope (GBT) to target ELM WD candidates with the largest minimum companion 
masses.  The results from semesters GBT/05C-041, GBT/06A-051, GBT/07C-072, and 
GBT/10A-046 are previously published \citep{agueros09b, agueros09a, kilic13}.  
Here, we report the results for the 20 candidates observed in the semesters 
GBT/12A-431 and GBT/14A-438.  The 20 candidates were published in 
previous ELM Survey papers, but their radio observations were not.

	We selected targets for radio observations based on the minimum companion 
mass derived from the spectroscopic radial velocity curve and a Bayesian model 
estimate for the likelihood any particular system contains a neutron star companion 
\citep{andrews14}.  We obtained GBT observations with the GUPPI backend with a 
central frequency of 340 MHz, using 4096 channels each with 100 MHz bandwidth.  We 
set integration times to reach a detection threshold of 0.4 mJy kpc$^2$.  Data were 
processed, de-dispersed, and folded using standard routines within {\tt PRESTO} 
\citep{ransom02, 
ransom03}\footnote{\href{https://www.cv.nrao.edu/~sransom/presto/}{https://www.cv.nrao.edu/$\sim$sransom/presto/}}. 
Our procedure searched dispersion measures as large as twice the expectation from 
the spectroscopic distances, using the Galactic electron density model from 
\citet{cordes02}.

	The result is a single new pulsar, PSR J0802$-$0955.  However, follow-up 
radio observations \citep{andrews18} indicate this pulsar is either a foreground or 
background object, unrelated to the coincident ELM WD.  We therefore identify no 
radio pulsar companions in the ELM Survey.

	Null detections place useful lower limits on inclination if we assume the 
secondaries have $M_2<1.4$~\msun.  We list the GBT targets and their inclination 
constraints in the Appendix.

\subsubsection{X-ray Observations}

	We obtained {\it Chandra} X-ray Observatory \citep{weisskopf02} observations 
for ten low mass ELM candidates.  Eight are previously published \citep{kilic13, 
kilic14b, kilic16}.  We present the results for the final two candidates with 
{\it Chandra} observations, J0147+0113 and J2245+0750 here.

	We observed J2245+0750 for 123 kiloseconds on 2018 September 17-22, and 
J0147+0113 for 13.9 kiloseconds on 2018 November 11.  We placed ACIS-S at the focus, 
in Very Faint mode. No periods of enhanced background were seen, so we extracted 
spectra from 1.5\arcsec radii at the location of each WD, and fit them with a 
hydrogen neutron star atmosphere model \citep{heinke06}. We assumed a distance of 
1.5 kpc for J2245+0750 and 810 pc for J0147+0113 from our spectroscopic distance 
estimates.  We assume $N_H=6.35\times10^{20}$ cm$^{-2}$ for J2245+0750 and 
$N_H$=$2.87\times10^{20}$ cm$^{-2}$ for J0147+0113, from the \citet{dickey90} 
reddening estimates in these directions. We fix the assumed neutron star mass and 
radius to 1.4 \msun\ and 12~km, respectively, and the temperature to $\log{T} 
=5.903$, the lowest observed for any millisecond pulsar in 47~Tuc 
\citep{bogdanov06}, with the normalization (thus the area of the hot spot) free.

	We detect no sources at the location of either WD.  We thus fit the spectra 
using the C-statistic \citep{cash76}, and obtain upper limits on the normalization, 
and thus on the X-ray luminosity. We find 90\% confidence limits of $L_X$(0.3-8 
keV)$\le2.6\times10^{29}$ erg~s$^{-1}$ for J2245+0750, and $\le2.2\times10^{29}$ 
erg~s$^{-1}$ for J0147+0113. These values are below the X-ray luminosities of any 
well-measured millisecond pulsars \citep{bogdanov06, kargaltsev12, forestell14}, 
allowing us to confidently rule out a millisecond pulsar companion in both cases.  
The unseen companions are likely WDs.  

	We can again use the null detections to place lower limits on inclination.  
We list the {\it Chandra} targets and their inclination constraints in the Appendix.

 \begin{figure} 	
 \includegraphics[angle=270,width=3.25in]{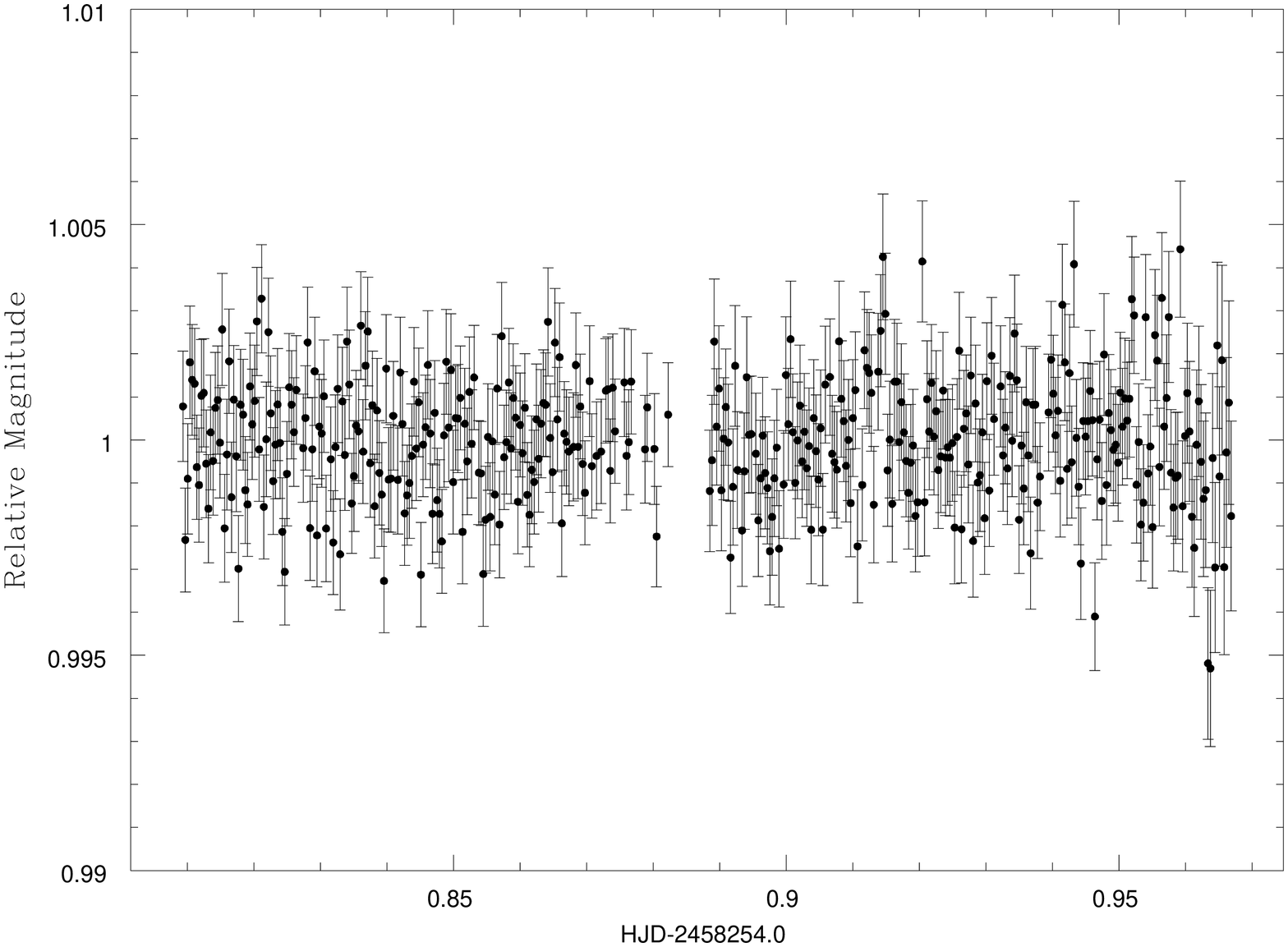} 
 \includegraphics[angle=270,width=3.25in]{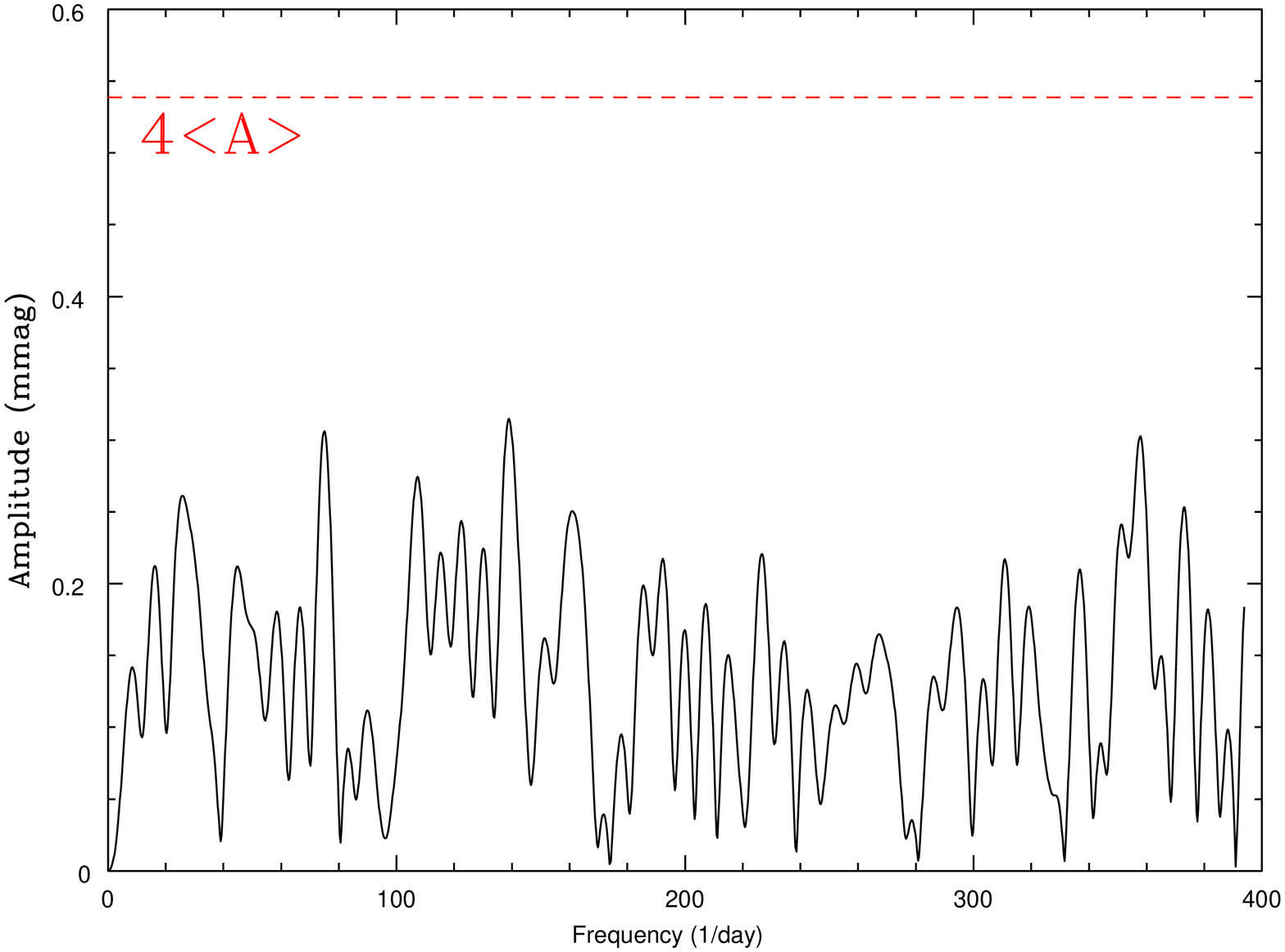} 
 \caption{ \label{fig:1738lc}
	Optical light curve of J1738+2927 (upper panel) and Fourier transform of the 
light curve (lower panel).  The gap in the light curve is due to zenith-crossing.  
The Fourier transform shows no evidence for significant variability.
 } \end{figure}

\subsection{Optical Light Curve}

	Finally, we obtained time-series optical photometry for the newly 
discovered $P=0.048$~d WD binary J1738+2927 on UT 2018 May 16.  Our goal was 
to check for eclipses. We acquired images using the Agile frame-transfer camera and 
the BG40 filter on the 3.5-meter telescope at the Apache Point Observatory (APO).  
We obtained 402 $\times$ 30~s exposures over a time baseline of 3.8~h (3.3 orbital 
periods) with median seeing of 1.16$\arcsec$ and airmass ranging from 1.0 to 1.1. 
Since J1738+2927 passes almost overhead at APO, we could not observe it for about 
20~min when it was near zenith, causing a small gap in coverage.

	Figure~\ref{fig:1738lc} shows the light curve for J1738+2927 and its Fourier 
transform. There are no significant peaks above the 4$<$A$>$=0.7\% level, suggesting 
that there is no significant variability in the light curve.  We check for Doppler 
boosting using equations 3 and 4 of \citet{shporer10}.  Based on the velocity 
semi-amplitude and minimum companion mass, we estimate the maximum magnitude of 
relativistic Doppler boosting to be $3.4\pm0.1\times10^{-3}$, or about 0.3\%. The 
predicted signal is undetectable given our measurement uncertainties.  The absence 
of eclipses implies the binary inclination is $i<85.8\arcdeg$.

\section{Discussion}

	Because the ELM Survey is selected on the basis of magnitude, color 
(temperature), and surface gravity, we can fairly test unrelated parameters like 
binary fraction and orbital period.  Orbital period and WD mass provide fundamental 
links to evolutionary models, binary population synthesis models, and future 
gravitational wave measurements.

\subsection{ELM WD Binary Fraction}

	Our observations are consistent with 100\% of ELM WDs being binaries.  
Quantitatively, 95\% (59/62) of the clean ELM WD sample are binaries with 
significant radial velocity orbital motion.  We do not expect to detect radial 
velocity motion in binaries with $i<20\arcdeg$ \citep{brown16a}.  Forward-modeling 
mock sets of binaries with the observed period and inferred companion mass 
distributions \citep{andrews14}, we estimate that 8\% (5/62) of simulated ELM WD 
binaries should not appear significantly velocity variable to our measurements.  
Observing 3 non-variable ELM WDs in the clean sample is thus statistically 
consistent with the number of face-on binaries we expect in a set of 62 randomly 
inclined binaries.  The previously reported excess of non-variable ELM WD candidates 
\citep{brown16a} is explained by mis-identified subdwarf A-type stars contaminating 
the ELM Survey at $<9,$000~K.

	Higher mass WDs have a much lower binary fraction \citep{brown11}.  In the
SPY survey, the multiplicity of $M>0.45$ \msun\ WDs is 4\% (23/567)  
\citep{napiwotzki19}, or about 25$\times$ lower than the $M<0.3$ \msun\ WDs observed
here.  The distribution of periods is also expected to differ with mass
\citep[e.g.][]{lamberts19}.  


\subsection{Orbital Period Distribution}

\begin{figure} 
 \includegraphics[width=3.5in]{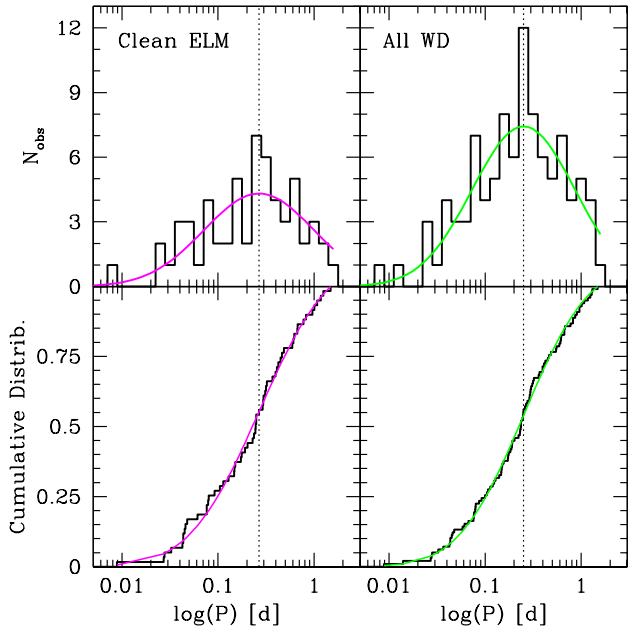} 
 \caption{ \label{fig:pdist}
	Period distribution of WD binaries in the clean ELM sample (left panels) and 
the full survey (right panels) with the best-fit lognormal distributions 
(solid lines).  Vertical dotted lines mark the lognormal means, $P\simeq0.25$~d.}
 \end{figure}

	The observed orbital periods in the ELM Survey range $0.0089<P<1.5$~d and 
are well-described by a lognormal distribution.  Figure~\ref{fig:pdist} plots the 
period distribution for the 59 binaries in the clean ELM WD sample (left) and the 98 
WD binaries in the entire survey (right).  Dotted lines mark the lognormal means, 
which are very near $P=0.25$~d.  The best fit parameters are lognormal $(\mu, 
\sigma)_{\rm Clean}=(-1.32, 1.32)$ and $(\mu, \sigma)_{\rm WD}=(-1.38, 1.23)$~d.

	The population of WD+dM binaries provide an interesting comparison 
\citep{rebassa07, rebassa10, rebassa12}.  WD+dM binaries have gone through a single 
phase of common envelope evolution, unlike the WD+WD binaries studied here, and are 
observed to have a wider range of periods $0.08<P<4.4$~d \citep{gomezmoran11}.  The 
orbital period distribution is linked to the common envelope ejection efficiency 
parameter \citep{zorotovic10}.  The longest-period binary is a constraint 
on the models \citep{li19}.

	Integrating our lognormal distributions to $P=\infty$ suggests there should 
be 10\% (about 6) more binaries in the clean ELM WD sample with $P>1.5$~d.  As seen 
in Figure~\ref{fig:pk}, the median companion in a $P=3$~d, $i=60\arcdeg$ orbit will 
result in $k=100$~\kms.  Yet we observe no $P=3$~d system.  J1021+0543 and 
J0802$-$0955 have the longest observed periods ($P=1.25$~d) with no aliases.  
To better constrain the long-period tail of the distribution will require higher 
precision measurements and/or longer observational time-baselines, i.e. for objects 
like J1512+2615, a $P=1.5$~d ELM WD with significant aliases.

\subsection{Mass-Period Distribution: Link to Formation}

	According to binary evolution theory, ELM WDs can form from either a stable 
Roche lobe overflow channel or a common envelope channel \citep{li19}.  The $P>1$~yr 
WD+MS binaries containing an ELM pre-WD \citep{vos18} or ELM WD \citep{masuda19, 
vikrant19} demonstrate that other evolutionary pathways also exist.  For double WD 
binaries, the diagnostic parameters are ELM WD mass and orbital period, because mass 
and period should be tightly correlated in the Roche lobe overflow channel.

\begin{figure} 
 \includegraphics[width=3.5in]{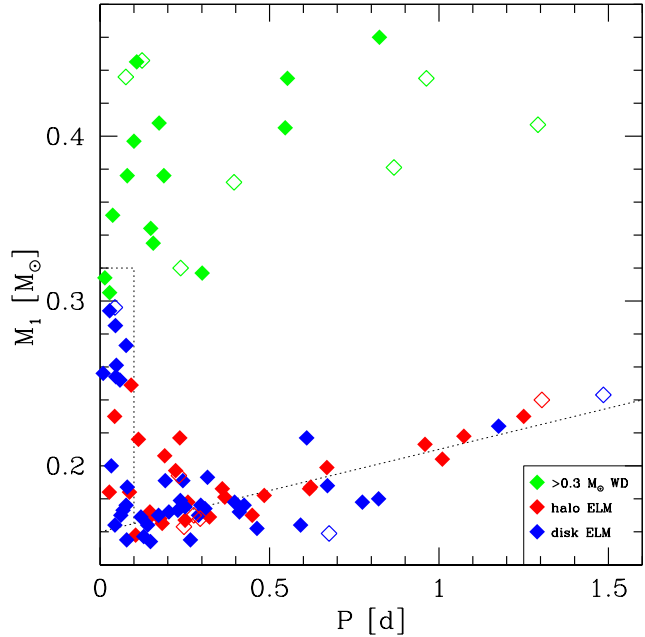}
 \caption{ \label{fig:mp}
	Mass of the visible WD versus binary period plotted on a linear scale.  
Blue and red points mark ELM WDs in the disk and halo, respectively.  All other WDs
with $>$0.3 \msun\ are plotted in green.  Binaries with period aliases are drawn
with a single open symbol at their best-fit period.  Dotted lines are drawn as a
guide; the distribution suggests that ELM WDs form from both Roche lobe overflow and
common envelope channels.}
 \end{figure}

	Following \citet{li19}, we plot mass versus orbital period for our updated
sample in Figure~\ref{fig:mp}.  Blue and red points are ELM WDs in the disk and
halo, respectively.  Green points are all the other WDs in the sample.  The major
uncertainty in this plot is systematic:  objects with period aliases, which are
drawn with open symbols at the best-fit period.  As previously noted, disk and halo
ELM WDs appear evenly mixed in parameter space.  We draw dotted lines in
Figure~\ref{fig:mp} as a guide to discussion.

	The diagonal band of binaries at the bottom of Figure~\ref{fig:mp} is likely
explained by the stable Roche lobe overflow channel.  In the formation models, the
ELM WD progenitors begin mass transfer near the end of the main sequence and produce
ELM WD masses correlated with period extending up to $M=0.3$ \msun\ \citep{li19}.  
The highest mass ELM WD we observe in this band is $M=0.24$ \msun.

	Between about 0.22 \msun\ and 0.32 \msun\ there is a vertical band binaries
seen only with $P<0.1$~d.  This group of ELM WDs likely comes from the common
envelope channel.  In the formation models, the ELM WD progenitors begin mass
transfer near the base of the red giant branch and produce more massive $M>0.21$
\msun\ ELM WDs due to the energy required to eject the common envelope \citep{li19}.  
Interestingly, half (12/25) of the WDs in our sample with $P<0.1$~d are observed in
this mass range.  The median gravitational wave merger time of these binaries is
$\tau=10^8$~yr.

	Finally, we observe $M>0.3$ \msun\ WDs with a diverse range of $P$. Our 
sample is not complete at these masses, but the observed period distribution appears 
consistent with the common envelope efficiency $\alpha_{\rm CE}=0.5$ binary 
population synthesis models of \citet{li19}.

\subsection{He+CO Merger Rate: Link to Outcomes}

	Once formed, ELM WD binary orbits shrink due to gravitational wave 
radiation.  The gravitational wave merger timescale depends 
primarily on period, \begin{equation}
	\tau = 47925 \frac{(M_1 + M_2)^{1/3}}{M_1 M_2} P^{8/3} ~{\rm Myr}
	\end{equation} where the masses are in \msun, the period $P$ is in days, and 
the time $\tau$ is in Myr \citep{kraft62}.  For the clean ELM WD sample, $\tau$ 
ranges from 1~Myr (J0651+2844) to 700~Gyr (J1512+2615) and has a median value of 
$10$~Gyr.

	Physically, ELM WDs are He-core WDs.  Their unseen companions are typically 
0.75~\msun\ objects at 1.6~\rsun\ orbital separations -- thus CO-core WDs -- if the 
binaries have random inclination \citep{andrews14, boffin15, brown16a}.

	ELM WD binaries are thus He+CO WD binaries with typical mass ratios of about 
1:4.  A 1:4 mass ratio suggests that most binaries will evolve into stable helium 
mass-transfer systems, so-called AM CVn stars \citep{marsh04}.  However, the 
dynamically driven double-degenerate double-detonation scenario posits that 
essentially all He+CO WD binaries have unstable mass transfer \citep{shen15}.  We 
can test the outcome of He+CO WD mergers by comparing the merger rate against the 
formation rate of AM~CVn.

	We previously derived a merger rate for ELM WD binaries in the disk of the 
Milky Way using both reverse and forward-modeling approaches \citep{brown16b}.  The 
rate calculation is dominated by the shortest-period binaries.  The number of disk 
ELM WD binaries with $\tau<70$~Myr has grown by 33\% (from 6 to 8) with the addition 
of J1043+0551 \citep{brown17b} and J1738+2927 (this paper).  However, we now exclude 
J0935+4411 \citep{kilic14} from the clean ELM WD sample because the WD is 0.32 
\msun.  The completeness correction has also changed because follow-up is now 97\% 
complete in the clean ELM region.  The updated merger rate for ELM WD binaries in 
the disk of the Milky Way is $2\times10^{-3}$~yr$^{-1}$, 30\% lower than the 
previous estimate but consistent within its factor of 2 uncertainty.

	The 1:1 number ratio of binaries with $\tau<10$~Gyr and $\tau>10$~Gyr 
provides a complementary constraint.  Binaries with $\tau>10$~Gyr accumulate over 
time.  The only way to observe rapidly-merging systems without accumulating too many 
$\tau>10$~Gyr binaries is if the majority of ELM WD progenitors detach from the 
common envelope phase with $<$1~h orbital periods \citep{brown16b}.

	The upshot is that the merger rate of observed ELM WD binaries exceeds the 
formation rate of stable mass-transfer AM~CVn binaries in the Milky Way 
\citep{roelofs07b, carter13} by a factor of at least 25.  The total He+CO WD merger 
rate in the Galaxy can only be larger, because we do not observe all He+CO WDs.  
The ELM Survey observations thus require unstable mass transfer outcomes and support 
models in which most He+CO WDs merge \citep{shen15}.

\subsection{Gravitational Wave Sources}

	WD binaries with $P<1$~h emit gravitational waves at mHz frequencies and are 
potentially multi-messenger sources detectable by the future {\it LISA} 
gravitational wave observatory.  J0651+2844, for example, is an order of magnitude 
more luminous in gravitational waves (3~L$_{\odot}$) than in bolometric light 
(0.1~L$_{\odot}$).


	\citet{lamberts19} recently combine binary population synthesis models with 
cosmological simulations of Milky Way-like galaxies to predict what type of binaries 
{\it LISA} will see \citep[see also][]{korol17, breivik19}.  He+CO WD binaries, like 
those observed here, are predicted to be 50\% of the binaries individually resolved 
by {\it LISA}.  The majority of sources should be in the disk, though the bulge and 
halo are also predicted to contribute detections.

\begin{figure} 
 \includegraphics[width=3.5in]{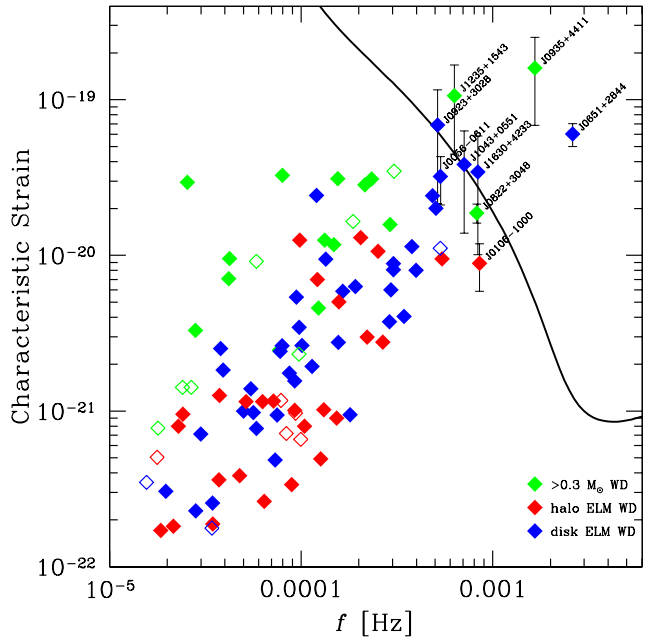}
 \caption{ \label{fig:gw}
	Characteristic strain $h_c$ times $\sqrt{(4~{\rm yr})f}$, the S/N boost from 
the number of cycles during the {\it LISA} observation time, versus gravitational 
wave frequency $f=2/P$ for all 98 WD binaries.  Symbols are the same as 
Figure~\ref{fig:mp}.  Solid line is the {\it LISA} 4~yr sensitivity curve 
\citep{robson19}.}
 \end{figure}

	To compare with the binaries we observe optically, Figure~\ref{fig:gw} plots 
the characteristic strain versus gravitational wave frequency $f=2/P$ for all 98 WD 
binaries.  We also draw the 4~yr {\it LISA} sensitivity curve (solid line) 
\citep{robson19} as a guide for discussion. Symbols and colors are the same as in 
Figure~\ref{fig:mp}.

	We compute characteristic strain using inclination,
	\begin{equation} h_c = 3.4\times10^{-23} \sqrt{\cos^4(i)+6\cos^2(i)+1} 
\mathcal{M}^{5/3} P^{-2/3} d^{-1}, \end{equation} where $\mathcal{M}= (M_1 
M_2)^{3/5} (M_1+M_2)^{-1/5}$ is the chirp mass in \msun, $P$ is in days, and $d$ is 
in kpc \citep{timpano06, roelofs07a}.  We multiply by $\sqrt{(4~{\rm yr})f}$ to 
account for the S/N boost from the number of cycles during the {\it LISA} 
observation time \citep{robson19}.  We note that most strain calculations implicitly 
assume $i=60\arcdeg$, which yields a strain systematically 1.6$\times$ too large for 
eclipsing binaries like J0651+2844.  Ironically, non-velocity-variable ELM WDs may 
be among the highest strain systems since they (presumably) have low inclination, 
however we have no constraints on their orbital periods.

	Our approach to Figure~\ref{fig:gw} is to compute strain 10,000 times per 
binary assuming random inclination and normally distributed measurement errors, 
including any inclination constraint.  Four WD binaries in the ELM Survey are 
eclipsing \citep{steinfadt10, brown11b, brown17b, kilic14}, eight have ellipsoidal 
variations \citep{kilic11b, hermes12a, hermes14, bell17}, and 32 have X-ray and/or 
radio observations that rule out low inclination millisecond pulsar companions 
\citep{agueros09a, agueros09b, kilic11a, kilic12a, kilic14b, kilic16, andrews18}.  
We also exclude inclinations that correspond to physically unlikely mass extremes:  
companions less massive than the observed ELM WD, or greater than 3 \msun.  We 
detail the inclination constraints in the Appendix.  Errorbars in 
Figure~\ref{fig:gw} are the 16\% and 84\% percentiles of the resulting strain 
distribution, and the values are listed in Table \ref{tab:orb}.  For the sake of 
clarity, we label only those WD binaries near the 4~yr {\it LISA} sensitivity curve.

	Interestingly, all six binaries on or above the 4~yr {\it LISA} sensitivity 
curve are disk objects, as predicted by the models.  The strongest halo binary, 
J0822+3048, falls just below the sensitivity curve.  However, the \citet{lamberts19} 
models do not match the observed distribution of periods.  The model period 
distribution has a gap around $f=0.8$ mHz where most (4 of 7) of the observed 
binaries with periods below 1~h reside in our sample.

	The highest S/N source J0651+2844.  Because its spectroscopic distance is 
many times more accurate than its {\it Gaia} DR2 parallax, its characteristic strain 
in Figure~\ref{fig:gw} has a much smaller uncertainty than calculated by 
\citet{kupfer18}.  According to the LISA Detectability Calculator, J0651+2844 has a 
4-yr S/N~$\simeq150$ (Q.~S.~Baghi, private communication).  It would be very 
interesting to find more WD binaries like J0651+2844.  Being a sample of one, 
however, implies that we may need to observe $\sim$100 more ELM WDs to find another 
J0651+2844.

	A more productive approach to finding strong mHz gravitational wave sources
may be to target bright and nearby ELM WD candidates.  In the clean ELM WD sample,
10/62 (or 1-in-6) of the binaries have $P<0.05$~d (or $f>0.5$~mHz) where LISA is
most sensitive.  An untargeted approach, taken by the Zwicky Transient Factory,
is to search for short-period eclipsing systems from all-sky time-series imaging
\citep{burdge19, burdge19b}.

\section{Conclusions}

	The ELM Survey was a major observational program that targeted extremely low 
mass, He-core WDs on the basis of magnitude and color.  It is now essentially 
complete within our color/magnitude selection limits in the SDSS footprint.  One of 
the major goals of this paper is to consolidate all the measurements:  4,338 radial 
velocity measurements, 230 stellar atmosphere fits, 128 radial velocity orbital 
solutions, and 47 inclination constraints derived from follow-up optical, X-ray, and 
radio observations.  New measurements include {\it Chandra} and GBT observations, 
plus stellar atmosphere fits and radial velocity solutions for 25 well-constrained 
new binaries.

	We apply {\it Gaia} parallax and proper motion measurements to the sample 
for the first time, and find that ELM WD evolutionary tracks provide accurate 
luminosity estimates for candidates with $T_{\rm eff}>9,$500~K.  However, most
candidates with $<$9,000~K have radii and luminosities too large to be WDs on 
the basis of their parallax, and so are subdwarf A-type stars (a.k.a.\ field blue 
stragglers).  This motivates us to define a clean set of WDs over which our 
observations are complete.

	The ELM Survey contains a total of 98 WD+WD binaries, more than half of
known detached double WD binaries in the Milky Way.  In the clean sample, 35\% of
the binaries are halo objects on the basis of 3D space motions.  Their orbital
periods span $0.0089 < P < 1.5$~d and are correlated with He WD mass, providing
evidence for both stable Roche lobe and unstable common envelope formation channels.  
We infer that most systems are He+CO WD binaries.  The gravitational wave merger
timescales imply a $2\times10^{-3}$~yr$^{-1}$ merger rate of He+CO WD binaries in
the disk of the Milky Way, which is a lower limit because we do not target all He+CO
WD binaries.  The merger rate is 25 times larger than the formation rate of stable
mass transfer AM CVn binaries, thus our observations require unstable mass transfer
outcomes for He+CO WD binary mergers \citep{shen15, brown16b}.

	The observed binaries notably emit gravitational waves at mHz frequencies.  
Two ELM Survey discoveries, J0651+2844 and J0935+4411, will be detected at high S/N 
by the future {\it LISA} mission.  Linking light and gravitational waves is 
important for making measurements beyond what either observational technique can 
achieve on its own.  Tidal dissipation, for example, is expected to significantly 
influence WD temperature and rotation prior to mass transfer and merger 
\citep{fuller14} and to appear as an accelerated $\dot{P}$ \citep{piro11, piro19}.  
Eclipse timing already provides exquisite $\dot{P}$ measurements for binaries like 
J0651+2844 \citep{hermes12c, hermes20}, ZTF J1539+5027 \citep{burdge19}, and PTF
J0533+0209 \citep{burdge19b}, however optical constraints on {\it mass} are much 
less precise.  {\it LISA} can provide an independent mass constraint for these 
systems, and, in conjunction with the optical $\dot{P}$, constrain the amount of 
tidal heating in merging pairs of WDs.

	It would thus be very interesting to find more WD binaries that can serve as
multi-messenger laboratories, systems we can observe with both light and gravity.  
To that end, we have begun the ELM Survey South \citep{kosakowski20}, targeting
southern hemisphere ELM WD candidates using photometric surveys like VST Atlas
\citep{shanks15} and SkyMapper \citep{wolf18}.  {\it Gaia} DR2 opens a new window on
target selection using parallax, which works well at bright $G<18.5$ magnitudes
\citep[e.g.][]{pelisoli19b}.  Over the past year, we have also observed new ELM WD
candidates using {\it Gaia}. Photometric surveys like the Zwicky Transient Facility,
and the Large Synoptic Survey Telescope will help immensely as well \citep{korol17}.  
The future of ELM WD discoveries appears bright both in light and gravitational
waves.\\

\acknowledgements

	We thank B. Kunk, E.\ Martin, and A.\ Milone for their assistance with 
observations obtained at the MMT Observatory, a joint facility of the Smithsonian 
Institution and the University of Arizona.  This research is based in part on 
observations obtained with the Apache Point Observatory 3.5-meter telescope, which 
is owned and operated by the Astrophysical Research Consortium.  This project also 
makes use of data obtained at the Southern Astrophysical Research (SOAR) telescope, 
which is a joint project of the Minist\'{e}rio da Ci\^{e}ncia, Tecnologia, 
Inova\c{c}\~{a}os e Comunica\c{c}\~{a}oes do Brasil, the U.S.\ National Optical 
Astronomy Observatory, the University of North Carolina at Chapel Hill, and Michigan 
State University.  This research makes use the SAO/NASA Astrophysics Data System 
Bibliographic Service.  This project makes use of data products from the Sloan 
Digital Sky Survey, which is managed by the Astrophysical Research Consortium for 
the Participating Institutions.
	This work was supported in part by the Smithsonian Institution, and in part 
by the NSF under grant AST-1906379. Additional support for this work was provided by 
NASA through Chandra Award Number G09-20021X issued by the Chandra X-ray Center, 
which is operated by the Smithsonian Astrophysical Observatory for and on behalf of 
NASA under contract NAS8-03060.  COH was supported by NSERC Discovery Grant 
RGPIN-2016-04602 and a Discovery Accelerator Supplement.

\facilities{MMT (Blue Channel spectrograph), FLWO:1.5m (FAST spectrograph), SOAR 
(Goodman spectrograph), Gemini (GMOS spectrograph), Mayall (KOSMOS), APO 
(Agile), CXO, GBT}

\software{IRAF \citep{tody86, tody93}, ~RVSAO \citep{kurtz98},  ~PRESTO 
\citep{ransom02, ransom03},  ~CIAO \citep{fruscione06}}

\appendix \section{WD Binary Inclination Constraints}

	Table~\ref{tab:i} summarizes the inclination constraints for WD binaries in 
the ELM Survey, with links to the papers that published the measurements.  The best 
inclination constraints come from time-series optical photometry 
\citep[e.g.][]{hermes14}.  Four eclipsing binaries (labeled Eclip.=1) have 
$i\simeq90\arcdeg$ with $\sim$1$\arcdeg$ uncertainties.  Ellipsoidal variation 
caused by tidal deformation of the WD also places an inclination constraint.  Eight 
binaries with ellipsoidal variation (labeled E.V.=1) have $i=50-75\arcdeg$ with 
$\sim$10$\arcdeg$ uncertainties \citep{bell18b}.  The {\it absence} of eclipses 
places another, weak, $i\lesssim88\arcdeg$ constraint for WD binaries with 
well-measured optical light curves.

	Radio and X-ray null detections place lower limits on inclination.  As 
mentioned above, milli-second pulsars are commonly observed with low mass WD 
companions \citep{manchester05, panei07}.  In all cases, however, ELM WD binaries 
with targeted {\it Chandra} (labeled X-ray=1) or Green Bank Telescope (labeled 
Radio=1) observations capable of detecting plausible pulsar companions find null 
detections.  Null detections imply $M_2<1.4$ \msun.  Solving the binary mass 
function for inclination,
	\begin{equation} i = {\rm asin} \left( k \left( \frac{P}{2\pi G} \right)
^{1/3} \frac{(M_1+M_2)^{2/3}}{M_2} \right) ,
	\end{equation} an upper limit on $M_2$ places a lower limit on $i$ 
given the binary's observed semi-amplitude $k$, period $P$, and derived mass $M_1$.

	An optional inclination constraint, not listed in Table~\ref{tab:i}, is the 
upper limit that comes from requiring $M_2>M_1$.  Because the most massive star in a 
binary should evolve first, it is implausible for ELM WDs to have companions of 
lower mass.  In practice, requiring $M_2>M_1$ provides only a weak inclination 
constraint; it affects the five ELM WD binaries in the clean sample with minimum 
$M_2$ less than the ELM WD mass (see Tables~\ref{tab:param} and \ref{tab:orb}).


\begin{deluxetable}{ccccccl}   
 \tabletypesize{\footnotesize} \tablecolumns{7} \tablewidth{0pt} 
 \tablecaption{ELM Survey WD Binary Inclination Constraints\label{tab:i}} 
 \tablehead{
   \colhead{Object} & \colhead{$i$} & \colhead{Eclip.} & \colhead{E.V.} 
	& \colhead{X-ray} & \colhead{Radio} & {Reference} \\
	& (deg) & & & & & }
	\startdata 
J0022$+$0031 & $               >21$ & 0 & 0 & 0 & 1 & 2 \\
J0022$-$1014 & $         >18,<86.0$ & 0 & 0 & 0 & 1 & 2,9 \\
J0056$-$0611 & $  >37,50^{+9}_{-7}$ & 0 & 1 & 0 & 1 & 0,9,6 \\
J0106$-$1000 & $     56^{+11}_{-8}$ & 0 & 1 & 0 & 0 & 12,9,6 \\
J0112$+$1835 & $ >45,66^{+10}_{-9}$ & 0 & 1 & 0 & 1 & 0,9,6 \\
J0152$+$0749 & $               >43$ & 0 & 0 & 0 & 1 & 0 \\
J0345$+$1748 & $      89.67\pm0.12$ & 1 & 0 & 0 & 0 & 11 \\
J0651$+$2844 & $        86.3\pm1.0$ & 1 & 1 & 0 & 0 & 10 \\
J0751$-$0141 & $>60,85.4^{+4.2}_{-9.4}$ & 1 & 1 & 1 & 1 & 0,17 \\
J0755$+$4800 & $         >52,<89.4$ & 0 & 0 & 1 & 1 & 0,18 \\
J0802$-$0955 & $               >71$ & 0 & 0 & 0 & 1 & 3 \\
J0811$+$0225 & $         >70,<88.4$ & 0 & 0 & 1 & 1 & 0,18 \\
J0822$+$2753 & $               >52$ & 0 & 0 & 1 & 1 & 2,14 \\
J0822$+$3048 & $88.1^{+1.4}_{-2.3}$ & 1 & 0 & 0 & 0 & 8 \\
J0825$+$1152 & $             <84.8$ & 0 & 0 & 0 & 0 & 9 \\
J0849$+$0445 & $         >45,<85.7$ & 0 & 0 & 1 & 1 & 2,14,9 \\
J0917$+$4638 & $               >29$ & 0 & 0 & 1 & 1 & 1 \\
J0923$+$3028 & $             <84.6$ & 0 & 0 & 0 & 0 & 9 \\
J0935$+$4411 & $               <70$ & 0 & 0 & 0 & 0 & 16 \\
J1005$+$0542 & $               >35$ & 0 & 0 & 0 & 1 & 0 \\
J1043$+$0551 & $             <85.7$ & 0 & 0 & 0 & 0 & 7 \\
J1053$+$5200 & $         >26,<82.0$ & 0 & 0 & 0 & 1 & 2,9 \\
J1054$-$2121 & $     72^{+9}_{-10}$ & 0 & 1 & 0 & 0 & 4,6 \\
J1056$+$6536 & $         >25,<84.9$ & 0 & 0 & 0 & 1 & 2,9 \\
J1104$+$0918 & $               >35$ & 0 & 0 & 0 & 1 & 0 \\
J1108$+$1512 & $             <87.2$ & 0 & 0 & 0 & 0 & 4 \\
J1141$+$3850 & $               >49$ & 0 & 0 & 0 & 1 & 0 \\
J1151$+$5858 & $               >46$ & 0 & 0 & 0 & 1 & 0 \\
J1233$+$1602 & $         >54,<90.0$ & 0 & 0 & 0 & 1 & 0,9 \\
J1234$-$0228 & $         >13,<71.5$ & 0 & 0 & 0 & 1 & 2,9 \\
J1238$+$1946 & $               >48$ & 0 & 0 & 0 & 1 & 0 \\
J1436$+$5010 & $         >36,<84.4$ & 0 & 0 & 0 & 1 & 2,9 \\
J1443$+$1509 & $         >55,<88.3$ & 0 & 0 & 1 & 1 & 0,18 \\
J1449$+$1717 & $             <87.6$ & 0 & 0 & 0 & 0 & 4 \\
J1518$+$0658 & $               >43$ & 0 & 0 & 0 & 1 & 0 \\
J1538$+$0252 & $               >42$ & 0 & 0 & 0 & 1 & 0 \\
J1618$+$3854 & $             <88.1$ & 0 & 0 & 0 & 0 & 5 \\
J1625$+$3632 & $               >10$ & 0 & 0 & 0 & 1 & 2 \\
J1630$+$4233 & $         >24,<82.8$ & 0 & 0 & 0 & 1 & 2,13 \\
J1738$+$2927 & $             <85.8$ & 0 & 0 & 0 & 0 & 0 \\
J1741$+$6526 & $  >64,75^{+7}_{-8}$ & 0 & 1 & 1 & 1 & 0,9,17,6 \\
J1840$+$6423 & $               >52$ & 0 & 0 & 0 & 1 & 0 \\
J2103$-$0027 & $               >48$ & 0 & 0 & 0 & 1 & 0 \\
J2119$-$0018 & $    68^{+10}_{-11}$ & 0 & 1 & 0 & 0 & 9 \\
J2132$+$0754 & $         >58,<87.5$ & 0 & 0 & 1 & 1 & 0,18 \\
J2236$+$2232 & $               >34$ & 0 & 0 & 1 & 1 & 15 \\
J2338$-$2052 & $             <83.8$ & 0 & 0 & 0 & 0 & 9 \\
	\enddata
 \tablerefs{ (0) this paper,
	(1) \citet{agueros09a}, (2) \citet{agueros09b}, (3) \citet{andrews18},
	(4) \citet{bell17}, (5) \citet{bell18}, (6) \citet{bell18b}, (7) \citet{brown17a}, 
	(8) \citet{brown17b}, (9) \citet{hermes14}, (10) \citet{hermes20},
	(11) \citet{kaplan14}, (12) \citet{kilic11b}, (13) \citet{kilic11c},
	(14) \citet{kilic12a}, (15) \citet{kilic13}, (16) \citet{kilic14}, 
	(17) \citet{kilic14b}, (18) \citet{kilic17}
	}
 \end{deluxetable}

 \clearpage


\end{document}